\documentclass[a4paper,fleqn,usenatbib]{mnras}

\usepackage[T1]{fontenc}
\usepackage{ae,aecompl}

\usepackage{amsmath}
\usepackage{amsfonts}
\usepackage{amssymb}
\usepackage{graphicx}
\usepackage{multirow,bigdelim}
\usepackage{adjustbox}
\usepackage{placeins}
\usepackage{soul}
\usepackage{array}

\newcolumntype{C}[1]{>{\centering\arraybackslash}m{#1}}

\def\CN2{\mbox{$C_N^2 \ $}}

\def\CT2{\mbox{$C_T^2 \ $}}

\def\sigmal2{\mbox{$\sigma ^{2}_{I} \ $}}

\title[Forecasting PWV above the sites of VLT and LBT]{Forecasting water vapour above the sites of ESO's Very Large Telescope (VLT) and the Large Binocular Telescope (LBT)}

\author[A. Turchi et al.]{
Alessio Turchi,$^{1}$\thanks{E-mail: aturchi@arcetri.astro.it}
Elena Masciadri,$^{1}$\thanks{E-mail: masciadri@arcetri.astro.it}
Florian Kerber,$^{2}$
Gianluca Martelloni$^{1,3}$
\\
$^{1}$INAF - Osservatorio Astrofisico di Arcetri, L.go E. Fermi 5, 50125  Florence, Italy\\
$^{2}$European Southern Observatory, Karl-Schwarzschild-Str.2, 85748 Garching, Germany\\
$^{3}$INSTM, Via della Lastruccia 3-13, 50019 Sesto Fiorentino (Florence), Italy\\
}

\date{Accepted 2018/09/28. Received YYY; in original form ZZZ}

\pubyear{2018}

\begin{document}
\label{firstpage}
\pagerange{\pageref{firstpage}--\pageref{lastpage}}
\maketitle

\begin{abstract}
Water vapour in the atmosphere is the main source of the atmospheric opacity in the infrared and sub-millimetric regimes and its value plays a critical role in observations done with instruments working at these wavelengths on ground-based telescopes. The scheduling of scientific observational programs with instruments such as the VLT Imager and Spectrometer for mid Infrared (VISIR) at Cerro Paranal and the Large Binocular Telescope Interferometer (LBTI) at Mount Graham would definitely benefit from the abitily to forecast the atmospheric water vapour content. In this contribution we present a study aiming at validating the performance of the non-hydrostatic mesoscale Meso-NH model in reliably predicting precipitable water vapour (PWV) above the two sites. For the VLT case we use, as a reference, measurements done with a Low Humidity and Temperature PROfiling radiometer (LHATPRO) that, since a few years, is operating routinely at the VLT. LHATPRO has been extensively validated on previous studies. We obtain excellent performances on forecasts performed with this model, including for the extremely low values of the PWV ($\leq$ 1~mm). For the LBTI case we compare one solar year predictions obtained with the Meso-NH model with satellite estimates again obtaining an excellent agreement. This study represents a further step in validating outputs of atmospheric parameters forecasts from the ALTA Center, an operational and automatic forecast system conceived to support observations at LBT and LBTI.
% It represents a preliminary step for a similar forecast system conceived for the VLT that we are setting-up.
 %This is a simple template for authors to write new MNRAS papers.
%The abstract should briefly describe the aims, methods, and main results of the paper.
%It should be a single paragraph not more than 250 words (200 words for Letters).
%No references should appear in the abstract.
\end{abstract}


%%%%%%%%%%%%%%%%%%%%%%%%%%%%%%%%%%%%%%%%%%%%%%%%%%

%%%%%%%%%%%%%%%%% BODY OF PAPER %%%%%%%%%%%%%%%%%%

\section{Introduction}
\label{intro}
Astronomical observations in the infrared (IR) part of the spectrum, from ground-based telescopes, are mainly affected by the water vapor content\footnote{In some cases OH lines are more critical} of the atmosphere, which is the main source of opacity in IR and sub-millimeter domains. Ground-based observations at these wavelengths can be done only in transparent windows of the spectrum without water vapour absorption lines. Atmospheric transmission depends on the distribution of atmospheric components, such as H$_{2}$O, O$_{2}$, CO$_{2}$ and other molecules. Water vapour is highly variable in the atmosphere both in altitude, time and geographic location. Atmospheric water content stratification is usually expressed in absolute humidity (AH), typically in $g/m^3$, which represents the total mass of water per volume of air. For astronomic purposes the parameter of interest is the integral of AH over the whole vertical height, which is the Precipitable Water Vapor (PWV), usually expressed in~mm, which represents the total mass of water vapour in a column of unit cross-sectional area extending from the surface to the top of the atmosphere.

In order to identify sites with the minumum water vapour content in the atmosphere and preferably those with the minimum variability, a lot of effort has been spent in the past in searching for atmospherically dry sites for ground-based telescopes, typically at high elevations \citep{otarola,tremblin,giovanelli,saunders, sims}.
%or in the free atmosphere i.e. SOFIA at 12.5~km \citep{roellig}. 
An accurate forecast of PWV conditions, performed with numerical models, can allow observatories using service mode operations \citep{Primas} to plan observations that require very low PWV conditions while optimizing the use of available observing time. This will allow the possibility to plan in advance service-mode observations and match the science targets of a specific observing program with the corresponding optimal weather conditions, thus maximizing the scientific output of an observatory. 
A few studies already tried to apply mesoscale models to PWV forecast on astronomical sites \citep{giordano2013,pozo2016,perez2018}.

The main goal of this paper is to validate and quantify the reliability of the mesoscale atmospherical model Meso-NH \citep{lafore98,lac2018} in forecasting the PWV above Cerro Paranal, site of the Very Large Telescope (VLT) and Mt.Graham, site of the Large Binocular Telescope (LBT). 

VISIR \citep{lagage} is the instrument (imager and spectrometer) at VLT which operates in the mid infrared regime (5-20 $\mu$m) and it is strongly dependent on the PWV. Besides that, a few others instruments, running at shorter wavelengths (near-infrared or visible) benefit, at different levels, from the low levels of PWV and, in general, from knowing in advance the quantity of water present in the atmosphere. Among these, CRIRES \citep{kaeufl2004}, X-SHOOTER \citep{vernet2011}, MUSE \citep{henault2003} and at VLTI GRAVITY \citep{eisenhauer2008}. In the latter case the PWV fluctuations between the different telescopes, represent an important deterioration in interferometric observations. 

At LBT, the LBTI instrument \citep{hinz} (i.e. Large Binocular Telescope Interferometer) is running in the 2.9-13 $\mu$m range and it is highly affected by the PWV too. To appreciate the importance to forecast the exact value of the PWV content in an operational configuration we note that it has been observed that the VISIR sensitivity in Q band (17-25 $\mu$m) can allow to detect targets up to 3 mag higher in extremely low PWV conditions (0.1-0.2~mm) \citep{smette2008}. It is straightforward to conclude that a good managing of the PWV forecast may open new scientific opportunities both for imaging and spectroscopy from ground-based telescopes, as described e.g. in \citet{kerber2014}.

The interest for validating PWV provided by Meso-NH is twofold. From one side we have installed an operational and automatic system (ALTA Center\footnote{\url{http://alta.arcetri.astro.it/}}) conceived for the forecast of a set of atmospheric parameters relevant for the ground-based astronomy and astroclimatic parameters above Mt.Graham to support observations of LBT and LBTI \citep{masciadri1999,masciadri2017}. ALTA Center uses the Meso-NH and the Astro-Meso-NH codes. The first code is useful to predict atmospheric parameters, the latter to predict the optical turbulence ($\CN2$) and the integrated astroclimatic parameters (seeing, isoplanatic angle and wavefront coherence time). PWV is part of the atmospheric parameters and we intend to prove the model performances in forecasting the PWV that is extremely useful for LBTI. On the other side, the implementation of a similar operational and automatic system for Cerro Paranal, site of the VLT is on-going. Previous studies performed on Cerro Paranal and Mt.Graham have already proven excellent performances of the Meso-NH model in forecasting atmospheric parameters such as wind speed and direction, temperature and relative humidity at the ground level \citep{lascaux2013,lascaux2015,turchi2017} and above Cerro Paranal excellent performances in reconstructing the vertical profiles of the same atmospherical parameters on around 20~km \citep{masciadri2013}. This study can provide information on the model ability in forecasting the PWV that is critical for instruments such as VISIR, CRIRES, X-SHOOTER, MUSE, GRAVITY.

Cerro Paranal is characterized by extremely dry conditions, with ground relative humidity typically in the range 5\%-20\% and a median value of PWV$\sim$2.4~mm \citep{kerber2012}. Starting from 2011, it has been installed at the VLT a Low Humidity and Temperature Profiling microwave radiometer (LHATPRO) that routinely monitors the PWV in this location. LHATPRO has been previously extensively validated by \citet{kerber2012} and it has been specifically engineered for monitoring dry sites. We have no equivalent instrument above Mt.Graham. The strategy of the paper is therefore to prove that the model is able to reconstruct reliable PWV above Cerro Paranal using LHATPRO measurements as a reference. This will be done by comparing observations with model outputs on two rich statistical samples in 2013 and 2017. Once the model has been validated for Cerro Paranal we apply the same model to Mt.Graham. This can be done because the forecast of the PWV does not require a model calibration such as that for the optical turbulence in which some specific parameters of the turbulent energy scheme are tuned to optimize the model behaviour for a specific site. 
In Section \ref{sec:mtg} we will discuss why this assumption is justified.

In Section \ref{sec:mod_conf} we present the model configurations used for the present study. In Section \ref{sec:obs} we describe the LHATPRO instruments used for measuring PWV and AH and the criteria used for the samples selection. In Section \ref{sec:res} we describe the results of the model validation performed at Cerro Paranal (VLT site) on different years and conditions in statistical terms. In Section \ref{sec:compar} we provide a comparison between predictions obtained with Meso-NH model and the ECMWF General Circulation Model. In Section \ref{sec:resah} we compare the vertical distribution of the AH observed and reconstructed by the Meso-NH model. In Section \ref{sec:spec} we present a test case done on an extremely low PWV event in 2017. In Section \ref{sec:mtg} we perform a climatological comparison done on Mount Graham (LBT site) between PWV estimated from Meso-NH and PWV measured with GOES satellites measurements \citep{carrasco2017}. Finally, in Section \ref{concl} we present our conclusions.\\

%%%%%%%%%%%%%%%%%%%%%%%%%%%%%%%%%%%%%%%%%%%%%
\section{Model Configuration}
\label{sec:mod_conf} 

We use the atmospheric model called Meso-NH\footnote{\url{http://mesonh.aero.obs-mip.fr/mesonh/}} (hereafter MNH) to forecast the PWV \citep{lafore98,lac2018}. This is a non-hydrostatic mesoscale model that computes the evolution of weather parameters in a three-dimensional volume over a finite geographical area. Is uses a forward in time (FIT) numerical scheme to compute the hydrodynamic equations. The coordinate system is based on mercator projection, while the vertical levels use the Gal-Chen and Sommerville coordinate systems \citep{chen1975}. We consider wave-radiation open boundary conditions with Sommerfeld equation for the normal velocity components \citep{somm}.  
The model itself is based on an anelastic formulation of hydrodynamic equations, which allows for the filtering of acoustic waves. The simulations make use of a one-dimensional mixing length proposed by \citet{Bougeault89} with a one-dimensional 1.5 closure scheme \citep{Cuxart00}. We take into account the interaction between surface and atmosphere parameters with the ISBA (Interaction Soil Biosphere Atmosphere) scheme \citep{Noilhan89}. We use a Kessler microphysical scheme \citep{kessler} for the water. The radiation scheme used is the ECMWF one \citep{hogan}. The long wave radiation is computed following the Rapid Radiation Transfer Model \citep{mlawer}. In the short wave, the ECMWF version of RRTM \citep{morcrette} is used.\\

The VLT site is located at Cerro Paranal (24.62528 S, 70.40222 W) at an height of 2635~m above sea level, while LBT is located at Mount Graham (32.70131 N, 109.88906 W) at an height of 3191~m above sea level.\\
Meso-Nh simulations are fed with the initialization data provided by the European Centre for Medium Weather Forecasts (ECMWF), calculated with their General Circulation Model (GCM) extend on the whole globe. \\
Simulations cover the night time at the specific site, which is relevant for astronomical observations on each site. The date of each simulation in this paper is identified by the UT day "J'' in which the night starts. For the Cerro Paranal case, in this study we simulate 15 hours initializing the model at 18:00 UT of the day "J", forcing the model each 6 hours with data coming from the GCM of the ECMWF and we treat/analyse results in the interval [00:00 - 09:00] UT of day "J+1''.  This interval permits to fit the nighttime during the whole solar year. \\
For the Mount Graham case instead, we initialize the model at 00:00 UT of day "J'' (nighttime is always enclosed in the same UT day) with the same forcing scheme at intervals of 6 hours. As will be described and explained in Section \ref{sec:mtg}, in the case of Mt.Graham model outputs will not be compared to measurements obtained in situ at specific dates but we compare model outputs with satellite boutputs in climatologic terms. We evaluate therefore the model outputs as they are calculated in the ALTA Center during a full solar year.\\
We use a grid-nesting technique \citep{Stein00} that consists of using a set of different imbricated domains, described in Tables \ref{tab:resol} and \ref{tab:resol2}, with a digital elevation model (DEM, i.e. orography) extended on smaller and smaller surfaces having a progressively higher horizontal resolution. In this way, using the same vertical grid resolution, we achieve the highest horizontal resolution on the innermost domain extended on a limited surface around the summit to provide the best possible prediction at the specific site. Each domain is centered on the telescope coordinates.\\

\begin{table}
\caption{Horizontal resolution of each Meso-NH imbricated domain at Cerro Paranal (VLT).} 
\label{tab:resol}
\begin{center}       
\begin{tabular}{cccc} 
\hline
\rule[-1ex]{0pt}{3.5ex}  Domain & $\Delta$X (km) & Grid points & Domain size (km) \\
\hline
\rule[-1ex]{0pt}{3.5ex}  Domain 1 & 10 & 80x80 & 800x800 \\
\rule[-1ex]{0pt}{3.5ex}  Domain 2 & 2.5 & 64x64 & 160x160 \\
\rule[-1ex]{0pt}{3.5ex}  Domain 3 & 0.5 & 150x100 & 75x50 \\
\hline
\end{tabular}
\end{center}
\end{table}

\begin{table}
\caption{Horizontal resolution of each Meso-NH imbricated domain at Mount Graham (LBT).} 
\label{tab:resol2}
\begin{center}       
\begin{tabular}{cccc} 
\hline
\rule[-1ex]{0pt}{3.5ex}  Domain & $\Delta$X (km) & Grid points & Domain size (km) \\
\hline
\rule[-1ex]{0pt}{3.5ex}  Domain 1 & 10 & 80x80 & 800x800 \\
\rule[-1ex]{0pt}{3.5ex}  Domain 2 & 2.5 & 64x64 & 160x160 \\
\rule[-1ex]{0pt}{3.5ex}  Domain 3 & 0.5 & 120x120 & 60x60 \\
\hline
\end{tabular}
\end{center}
\end{table}
The DEM used for domains 1 and 2, on both sites, is the GTOPO\footnote{\url{https://lta.cr.usgs.gov/GTOPO30}}, with an intrinsic resolution of 1~km. In domains 3 of Cerro Paranal we use the ISTAR\footnote{This was bought by the ESO from the ISTAR Company (Sophia Antipolis, Nice, France). The method is based on two stereoscopic images of the same location taken at different angles, obtained by SPOT satellites.}, with an intrinsic resolution of 500~m (16 arcsec), while in the case of Mount Graham we use the SRTM90\footnote{\url{http://www.cgiar-csi.org/data/srtm-90m-digital-elevation-database-v4-1}} \citep{srtm}, with an intrinsic resolution of approximately 90~m (3 arcsec). Even if in principle the SRTM DEM is available also above Cerro Paranal, in previous studies we observed that model outputs are better correlated to measurements using the ISTAR DEM. We therefore preferred to implement this solution in our study. \\
In our configuration the grid-nesting allows a 2-way interaction between each couple of father and son domains i.e. each couple of contiguous domains. Under these conditions the atmospheric flow in the inner domains is the most realistic because the atmospheric flow inside each domain is in a constant thermodynamic equilibrium with the outer domain's flow.\\
In the Paranal case we use 62 vertical levels on each domain, starting from 5~m above ground level (a.g.l.), while in the Mount Graham case we use 54 vertical levels on each domain, with the first grid point equal to 20~m a.g.l.. In both cases the levels have a logarithmic stretching of 20\% up to 3.5~km a.g.l. From this point onward the model uses an almost constant vertical grid size of $\sim$ 600~m up to $\sim$23~km, which is the top level of our domain. The grid mesh deforms uniformly to adapt to the orography, so the actual size of the vertical levels can stretch in order to accommodate for the different ground level at each horizontal grid point. The different size of the first grid point (5 and 20 meters) is due to the fact that instruments providing atmospheric measurements (wind, temperature, relative humidity and pressure) and observations of the optical turbulence are located at different heights above the ground in the two sites (for example \citet{lascaux2013,lascaux2015,masciadri2017,turchi2017}). This same model is used for the PWV predictions as well as for the forecast of all the parameters that we have just mentioned.\\
The PWV value is provided by the Meso-NH model with a time sampling of two minutes of simulated time and it is calculated in the innermost domain having a horizontal resolution of 500~m.

%%%%%%%%%%%%%%%%%%%%%%%%%%%%%%%%%%%%%%%%%%%%%
\section{Measurements and Instrumentation}
\label{sec:obs}

The instrument used as a reference in this paper is the Low Humidity and Temperature Profiling microwave radiometer (LHATPRO) that has been installed at Cerro Paranal in 2011 \citep{kerber2012} and since then runs continuosly 24/7 providing measurements stored in the ESO archive\footnote{\url{http://www.eso.org/asm/ui/publicLog?name=Paranal}}. This instrument is completely automated and is manufactured by Radiometer Physics GmbH. It uses multiple microwave channels in the frequency bands of 183 GHz (H$_{2}$O) and 51-58 GHz (O$_{2}$) in order to retrieve, among others, the humidity and temperature profiles up to 10~km of altitude above the ground level. Measurements are taken on 39 vertical levels with a resolution that varies from 10~m at the ground level up to 1~km at the topmost height. For more detailed description see \citet{Rose}. As explained in \citet{kerber2012}, the 183 GHz line is extremely important because it allows to resolve the extremely low levels of PWV present on a dry site such as Paranal (median value is around 2.4~mm).\\
LHATPRO was validated for astronomical use in 2011 \citep{kerber2012} against radiosoundings, showing a good correlation with measurements, an accuracy of 0.1~mm and a precision of 0.03 mm) for the PWV. The instrument starts to saturate for PWV values above 20~mm, which are however very rare bad weather events at Paranal (typically below 15~mm). Absolute humidity (AH) vertical profiles were however never considered in previous ESO studies. The vertical distribution of the humidity is not particularly relevant for astronomical applications. However we investigate also this parameter in this paper mainly to better understand the model behaviour and quantifying its performances and eventually the space for improvements in the calculation of the PWV.

To investigate both PWV and AH we first selected a sample of 120 nights in 2013. ESO gave us access to the vertical distribution of AH in the period [2012-2014]. The 120 nights are uniformly distributed on the year 2013 (approximately one night every 3 days, 10 days on each month, blindly selected). The PWV retrieved from LHATPRO is sampled every 5 s, while AH profiles are sampled with a frequency of one minute. Meso-NH outputs are sampled every 2 minutes (both PWV and AH). Similarly to what has been already done in other similar studies \citep{lascaux2015,turchi2017}, in order to efficiently compare model outputs and observations for an operational application, we performed a moving average of data with a window of one hour on both data sets (forecasts and measurements) to remove the high frequencies. After this procedure, data are resampled on 20 minutes and then compared. This permit to put in evidence the trend of measurements and model outputs. In the rare case that no measurement is available we discard the simulated values.\\

Besides this analysis performed on 2013 data, we decided to test the model also on a statistical sampled related to a more recent period, successive to 2016 March i.e. the time in which the ECMWF GCM have been up-graded to a horizontal resolution of 9~km instead of 16~km. This test has the goal to investigate if the Meso-NH model performances might improve using more accurate initialization data. We selected therefore another sample of 120 nights uniformly distributed in 2017 (approximately one night every 3 days, blindly selected) and we performed a separate comparison observations vs. mesoscale model. LHATPRO PWV data in ESO archive were missing from most of the dates in the period 2017/06/11 - 2017/07/13, with some other sporadic case of missing data, however we were still able to select at least 10 days in each moth with usable data.  \\

Finally, as we will see in Section \ref{sec:res}, to investigate the most challenging case i.e. the model performances in forecasting the extremely low PWV values (PWV$\leq$1~mm) we selected a third sample. The criterium we used is to select all nights of the solar year 2017 in which the PWV is lower than 1 millimeter for at least 30$\%$ of the night. We counted 35 nights (see Table \ref{sample201735n}). This was necessary because we want to maximize the number of cases in which PWV is weaker than 1 mm. In the previous samples the criterium of uniformity selection over a whole solar year resulted in a scarcity of events with PWV $\leq$ 1~mm. The investigation of this sample permits us to have a more solid estimate of the model performances in reconstructing the lowest conditions of PWV.  Looking at the cumulative distribution of the PWV on the observations on the full year of 2013 and 2017 (Fig.\ref{fig:parcumdists} - bold line) we retrieve that the percentage of PWV $\leq$ 1~mm is of the order of 12\%. Even if the percentage is relatively small, it is important for ESO to optimize the use of this portion of time to retrieve the best advantage for VISIR.  \\

At Mount Graham (LBT) we do not have the availability of accurate on-site measurements of the PWV. In this case we compared the statistics obtained from the forecasts of the PWV produced by the ALTA project (i.e. the operational version of the model we are treating here - see Section \ref{intro}) in the period of almost one solar year [from 2016/09/21 to 2017/06/08] with exception of the months of July and August in which the telescope LBT is close because of monsoon season, with satellite data obtained by Carrasco et. al. \citep{carrasco2017} over the years 1993-1999, for a total of 58 months of observations.\\

While LHATPRO provides us both direct measure of PWV (in~mm) and a profile of (AH) (in g/m$^{3}$), Meso-NH model gives us the vertical profile of water vapor mixing ratio M (kg/kg), pressure P (Pascal) and temperature T (Kelvin degrees). We can however obtain PWV and AH from M with the appropriate relations \ref{eq1} and \ref{eq2}.
\begin{equation}
  PWV=-\frac{1}{g\rho_{H_2O}}\int_{P_0}^{P_{top}}M dP
  \label{eq1}
\end{equation}
\begin{equation}
  AH={10^3}M\frac{P}{T R_d}
  \label{eq2}
\end{equation}
PWV expressed in mm and AH in g/m$^{3}$. In the above equations $\rho_{H_2O}=10^3 kg/m^3$ is the water density, $g=9.81$ m$\mathbf s^{-2}$ is the standard gravity acceleration and R$_d$=287.05 J/(kg$\times$K) is the specific gas constant for dry air. We integrate between the ground level pressure $P_0$ and the top level ($\sim$20~km a.g.l) pressure $P_{top}$. We note that the water vapor scale height is in the range 1.5-2.5~km. Above the latter height the water content decrease drastically and is typically negligible above 10~km \citep{Querel2016}.\\

%%%%%%%%%%%%%%%%%%%%%%%%%%%%%%%%%%%%%%%%%%%%%
\section{Model Validation on Cerro Paranal}
\label{sec:res}

Simulations done with the Meso-NH model related to the three samples described in Section \ref{sec:obs} have been performed and results have been compared to LHATPRO measurements. 

\begin{figure*}
\centering
%\begin{adjustbox}{max width=\textwidth}
\includegraphics[width=0.6\textwidth]{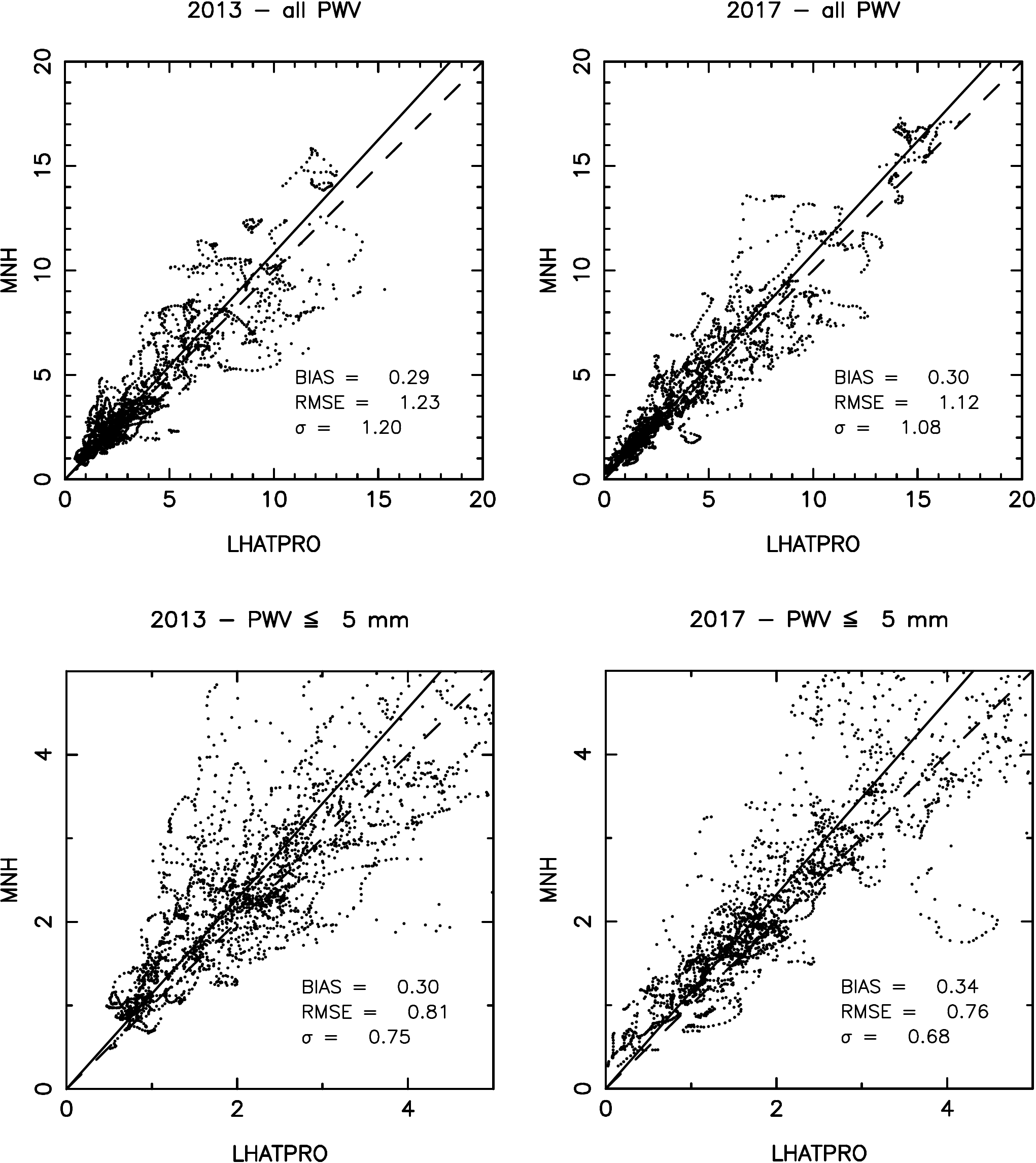}
%\end{adjustbox}
\caption{Cerro Paranal - Scatterplot computed on the raw data set PWV$_{raw}$. The plots are computed over the 120 nights 2013 sample (left column) and the 120 nights 2017 sample (right column). The indicators are computed over the whole sample (first row) and on the sub-sample in which PWV measurements are $\leq$ 5~mm (second row). The dashed line corresponds to the 45$^\circ$ that should represent a perfect match between model and measurements. The straight line corresponds to the regression line computed on the data points.}
\label{fig:rawscatter}
\end{figure*}

\begin{figure*}
\centering
%\begin{adjustbox}{max width=\textwidth}
\includegraphics[width=0.6\textwidth]{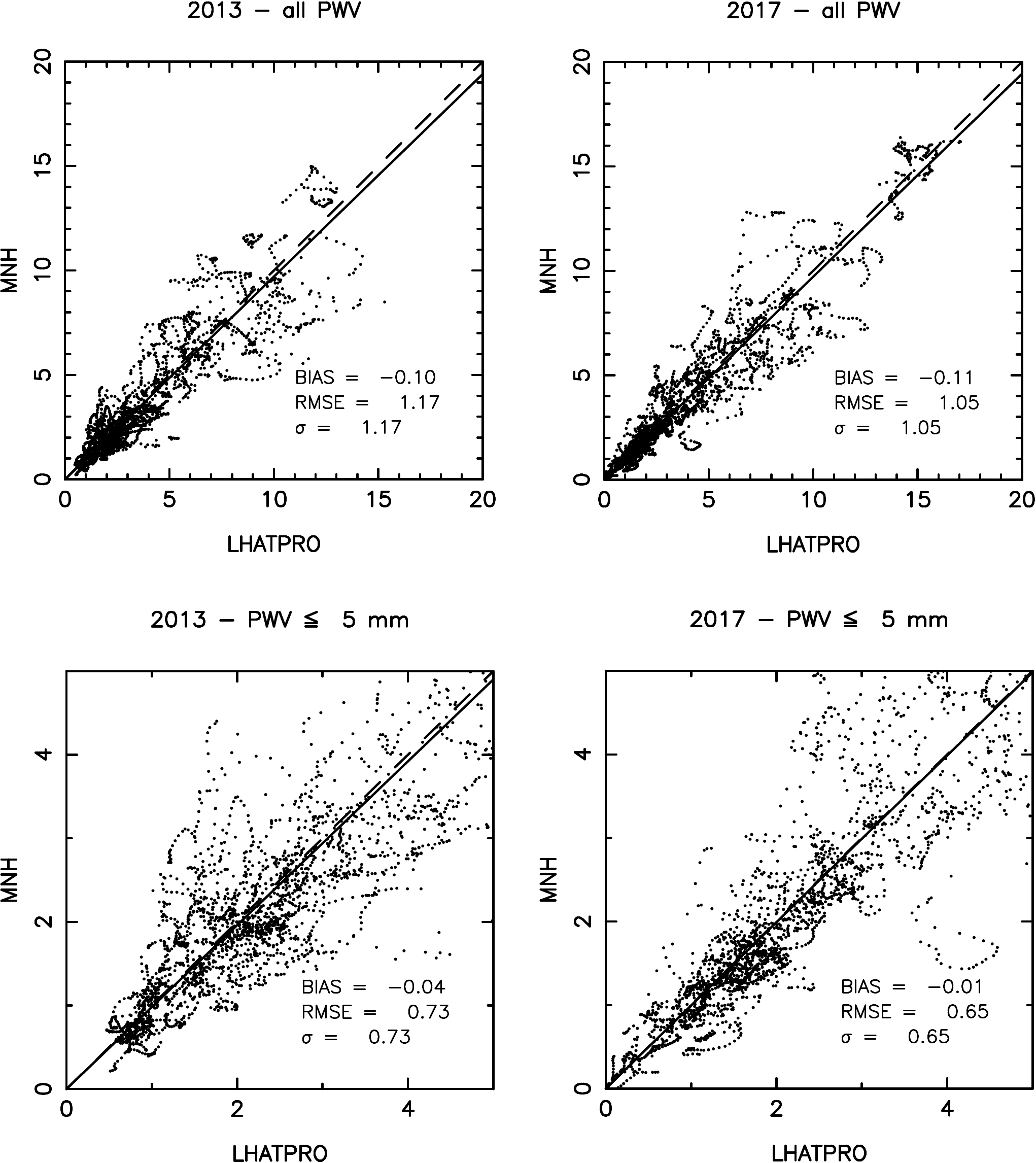}
%\end{adjustbox}
\caption{Cerro Paranal - Corrected data-set. Scatterplot computed on the corercted data set PWV$_{corr}$ (with correction in Eq. \ref{eq:regression}). The plots are computed over the 120 nights 2013 sample (left column) and the 120 nights 2017 sample (right column) . The indicators are computed over the whole sample (first row) and on the sub-sample in which PWV measurements with PWV$\leq$ 5~mm (second row). The dashed line corresponds to the 45$^\circ$ that should represent a perfect match between model and measurements. The straight line corresponds to the regression line computed on the data points.}
\label{fig:corrscatter}
\end{figure*}

%%%%%%%%%%%%%%%%%%%%%%%%%%%%%%%%%%%%%%%%%%%%%
\subsection{Model performances in reconstructing the PWV}
\label{sec:respwv}

A statistical analysis has been performed on the three independent samples of nights using the classical statistical operators BIAS, RMSE and $\sigma$, defined as:
\begin{equation}
BIAS = \sum\limits_{i = 1}^N {\frac{{(Y_i  - X_i )^{} }}
{N}} 
\label{eqbias}
\end{equation}
\begin{equation}RMSE = \sqrt {\sum\limits_{i = 1}^N {\frac{{(Y_i  - X_i )^2 }}
{N}} } 
\label{eqrmse}
\end{equation}
where $X_{i}$ are the individual observations and $Y_{i}$ the individual simulations computed at the same time index $i$, with $1\leq i\leq N$, $N$ being the total sample size. As done in previous studies \citep{lascaux2013,lascaux2015,turchi2017}, from the above quantities we deduce the bias-corrected RMSE ($\sigma$):
\begin{equation}\sigma = \sqrt {RMSE^2 - BIAS^2}
\label{eqsigma}
\end{equation}
The previously defined indicators, which provide us information on the statistical and systematic errors, were computed over two different data subsets. The first one is obtained by considering all the available measurements and forecasts while the second one is selected by considering only the $[X_{i}, Y_{i}]$ couples corresponding to LHATPRO measurements with PWV$\leq$ 5~mm. This allows us to properly characterize the model performance in low PWV value ranges, which are useful for telescope operations.\\

\begin{table}
 \begin{center}
 \caption{Cerro Paranal - In each column are the statistical indicators computed both on the raw model output and by applying correction in Eq. \ref{eq:regression}. The statistics are computed over the 120 nights 2013 sample (16 km resolution initialization data), the 60 nights 2017 sample (9 km resolution initialization data) and on a 60 nights subsample of the 2013 sample in order to make a comparison with the 2017 one unbiased by the sample size. The indicators are computed over the whole sample (All PWV) and on a sample filtered by selecting LHATPRO measurements with PWV$\leq$ 5~mm.} 
% \resizebox{\columnwidth}{!}{
 \begin{tabular}{c|ccc}
 \hline
 \multicolumn{1}{c}{} & RMSE (mm) & BIAS (mm) & $\sigma$ (mm) \\
 \multicolumn{1}{c}{} & (raw/corr) & (raw/corr) & (raw/corr) \\
 \hline
  {\bf 2013 - 120n} & &\\
   All PWV & 1.23/1.17 & 0.29/-0.10 & 1.20/1.17 \\
   PWV$\leq$ 5~mm &  0.81/0.73 & 0.30/-0.04 & 0.75/0.73 \\
   & & \\
 \hline
  {\bf 2017 - 120n} & &\\
   All PWV & 1.12/1.05 & 0.30/-0.11 & 1.08/1.05 \\
   PWV$\leq$ 5~mm &  0.76/0.65 & 0.34/-0.01 & 0.68/0.65 \\
   & & \\
 \hline
 \end{tabular}
% }
 \label{tab:rawstat}
 \end{center}
 \end{table}
 
We first analyzed the 120 nights 2013 sample. In Fig.\ref{fig:rawscatter}-left column we reported the corresponding scatterplots together with the computed regression lines (which must be compared to the dashed diagonal bisecting the plot). In Table \ref{tab:rawstat}-first row we reported the RMSE, BIAS and $\sigma$ computed in the previously specified ranges.
The same calculation is performed on the sample of 120 nights on 2017 (Fig.\ref{fig:rawscatter}-right column and Table \ref{tab:rawstat}-second row).

We notice that the statistical indicators computed on the 2013 and 2017 samples are consistent. Results are very convincing, since the dispersion of the data along the diagonal of the scatterplot is quite reduced. There is a cone effect with the dispersion increasing for large PWV values, confirmed by the statistical indicators computed on all the sample and on the PWV$\leq$ 5~mm range. In the latter case the RMSE is reduced by $\sim$1/3 in both samples, with respect to the full sample (from 1.23~mm to 0.81~mm in the 2013 sample and from 1.12~mm to 0.76~mm in the 2017 sample). The statistical operators RMSE and $\sigma$ are slightly better on 2017 with respect to 2013. The increased horizontal resolution of the initialization data passing from 2013 to 2017 can be the possible cause of this improvement. In both years (2013 and 2017), the RMSE in the PWV $\leq$ 5 mm case is well below the 1~mm. Results can be considered, therefore, very satisfactory.  
We also notice, in all cases, a residual BIAS of the order of $\sim$0.3~mm indicating that the model systematically slightly overestimates the PWV values measured by the LHATPRO. We refer to Section \ref{sec:resah} for a discussion on this point. This bias effect can however be reduced or eliminated with a regression model obtained from the statistics shown in the previous analysis. We searched for a regression which minimizes errors on all the datasets and on all the range of values of PWV and we selected an optimal correction reported in Eq. \ref{eq:regression}:
\begin{equation}
PWV_{corr} = \frac{PWV_{raw}}{1.04}-0.25
\label{eq:regression}
\end{equation}
where PWV$_{raw}$ is the uncorrected model output and PWV$_{corr}$ is the one corrected by the optimal regression line.\\

Table \ref{tab:rawstat} reports the PWV$_{corr}$ values obtained through Eq.\ref{eq:regression}, with the corresponding scatterplots shown in Fig. \ref{fig:corrscatter}. 

\begin{figure*}
\centering
%\begin{adjustbox}{max width=\textwidth}
\includegraphics[width=0.6\textwidth]{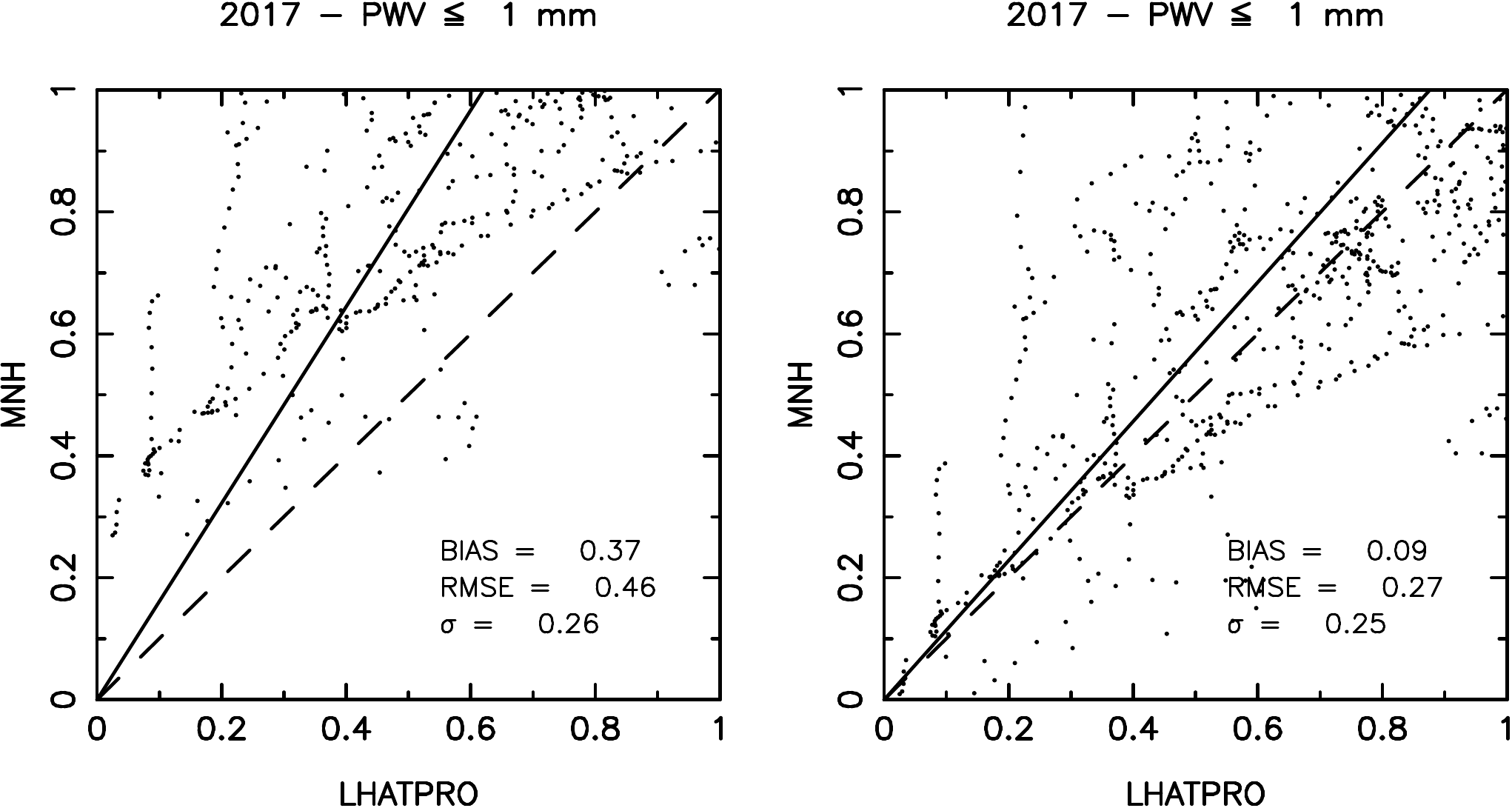}
%\end{adjustbox}
\caption{Cerro Paranal - Scatterplot computed on PWV. The plots are computed over the 35 nights 2017 sample obtained by selecting all nights on 2017 with PWV$\leq$ 1~mm for, at least, 30\% of the night. In left panel we show the uncorrected (raw) scatterplot, while in the right panel we show the scatterplot obtained with correction in Eq.\ref{eq:regression}. The indicators are computed by filtering the sample selecting LHATPRO measurements with PWV$\leq$ 1~mm. The dashed line corresponds to the 45$^\circ$ that should represent a perfect match between model and measurements. The straight line corresponds to the regression line computed on the data points.}
\label{fig:lowscatter}
\end{figure*}

We observe that the results are consistently improved on all the datasets, especially in the PWV$\leq$ 5~mm range. We are able to obtain a negligible BIAS on all the datasets and the RMSE is basically identical to $\sigma$, meaning that the residual error left is purely random, which confirms the validity of the applied correction. We also confirm that, concerning the PWV$\leq$ 5~mm dataset which is the most relevant one on Cerro Paranal (which have a median PWV$\sim$2.4~mm), the error on the 2017 dataset (RMSE=$\sigma$=0.65~mm) is lower than the one computed on the 2013 samples (RMSE=$\sigma$=0.73~mm). Since future initialization data from ECMWF will have a resolution which is equal or superior to the one available in 2017, we argue that we should consider the indicators computed on the 2017 sample as the reference for potential future applications. In this paper we do not study the correlation coefficient because in previous studies (\citet{masciadri2017} - Annex B) it has been shown that this parameter can be misleading. In many cases the model has an almost perfect agreement with measurements, however it shows a low correlation coefficient (due to small random fluctuations along the same baseline). In other cases the model may show a large bias or large local discrepancies with measurements, however it would have a good correlation coefficient. In other words, the correlation coefficient is not particularly relevant for the flexible-scheduling application. \\

Besides of the bias, RMSE and $\sigma$ we validated the model through the contingency table method, in a similar way to the previous studies on Cerro Paranal from \citet{lascaux2015}, which provide useful informations that complement the previous statistical analysis. While referring the reader to \citet{lascaux2015} for specific details on the method, here we report the main informations which are useful to understand the method. A contingency table is a method to analyse the relationship between variables in a categorical way. In practice we pre-define value intervals and we count the number of times in which the couples measurement/simulation $[X_{i}, Y_{i}]$ both fall into the same interval. With such defined tables it is possible to evaluate the probability of success of the model using several different statistical operators. As in previous studies we use the percentage of correct detection (PC), which represent the global probability of having both measurements and model to agree on the same interval. We also define the probability of detection (POD) in a specific range of values, which quantify the specific agreement probability for each defined range, and the extremely bad detection (EBD), which defines the probability of measurements and model to fall into distant categories (see definitions of PC, POD and EBD in \citet{lascaux2015} - Eq.9, 10,11,12 and 13).\\
If we define three range values (3$\times$3 table), in the case of a perfect model prediction we would have PC=PODs=100\% and EBD=0, while in the random prediction case we would have PC=PODs=33\% and EBD=22.5\%. It is also possible to define N$\times$N tables (see \citet{lascaux2015}) if it is desired. \\

Here we report the contingency tables computed on the full 2013 and 2017 samples after correction (see Eq. \ref{eq:regression}). We selected the following specified ranges: PWV $ \leq$ 1~mm, 1~mm $<$ PWV $\leq$ 5~mm, PWV $>$ 5~mm because this set appeared the most interesting from an observational point of view. Also 5~mm is approximately the third quartile of the PWV distribution over Cerro Paranal, as shown in figure \ref{fig:parcumdists}. In Tables \ref{tab:cont2013} and \ref{tab:cont2017} we see that model performance is satisfactory on both samples (2013 and 2017), with a PC=89.5\% on the 2013 sample and a PC=87.9\% on the 2017 sample. Specifically we note that we have EBD=0\% in both cases, meaning that the model never make dramatic errors. In each specified range the performance is comparable to the global PC. We note that the the most challenging probability of detection of PWV $\leq$ 1~mm (i.e. POD$_1$) presents excellent values for both 2013 (POD$_1$=81.8\%) and 2017 (POD$_1$=86.2\%). It is possible that the improved performance observed on 2017 with respect to 2013 is due to the higher horizontal resolution of the inizialization data. \\

 \begin{table}
 \begin{center}
 \caption{3$\times$3 contingency table - 2013 120 nights sample with applied correction in Eq. \ref{eq:regression}.} 
 %\resizebox{\columnwidth}{!}{
 \begin{tabular}{cc|ccc}
 \hline
 \multicolumn{2}{c}{PWV (mm)} & \multicolumn{3}{c}{\bf LHATPRO}\\
 \multicolumn{2}{c}{ } & PWV$\leq$      1 &      1 $<$PWV$\leq$      5  &  PWV$>$      5 \\
 \hline
  \multirow{7}{*}{\rotatebox{90}{\bf MNH}} & & &\\
   & PWV$\leq$         1 &           350      &           104      &             0     \\
   & & & & \\
   &          1 $<$PWV$\le$      5 &            78      &          1914      &            68     \\
   & & & & \\
   & PWV$>$         5 &             0      &            90      &           636     \\
   & & & & \\
 \hline
 \\
 \multicolumn{5}{l}{Sample size =       3240; PC=     89.5\%; EBD=      0.0\%} \\
 \multicolumn{5}{l}{POD$_1$=     81.8\%; POD$_2$=     90.8\%; POD$_3$=     90.3\%} \\
 \end{tabular}
 %}
 \label{tab:cont2013}
 \end{center}
 \end{table}

 \begin{table}
 \begin{center}
 \caption{3$\times$3 contingency table - 2017 120 nights sample with applied correction in Eq. \ref{eq:regression}.} 
 %\resizebox{\columnwidth}{!}{
 \begin{tabular}{cc|ccc}
 \hline
 \multicolumn{2}{c}{PWV (mm)} & \multicolumn{3}{c}{\bf LHATPRO}\\
 \multicolumn{2}{c}{ } & PWV$\leq$      1 &      1 $<$PWV$\leq$      5  &  PWV$>$      5 \\
 \hline
  \multirow{7}{*}{\rotatebox{90}{\bf MNH}} & & &\\ 
   & PWV$\leq$         1 &           154      &            57      &             0     \\
   & & & & \\
   &          1 $<$PWV$\leq$      5 &            16      &           832      &            35     \\
   & & & & \\
   & PWV$>$         5 &             0      &            20      &           349     \\
   & & & & \\
 \hline
 \\
 \multicolumn{5}{l}{Sample size =       3218; PC=     87.9\%; EBD=      0.0\%} \\
 \multicolumn{5}{l}{POD$_1$=     86.2\%; POD$_2$=     88.1\%; POD$_3$=     88.0\%}
 \end{tabular}
 %}
 \label{tab:cont2017}
 \end{center}
 \end{table}

If we look at the cumulative distribution calculated for LHATPRO and Meso-NH on the samples of 120 nights on 2013 and 2017 (Fig.\ref{fig:parcumdists}, black dashed line and red line) it is possible to conclude that the samples are perfectly representative of the typical conditions of PWV at Cerro Paranal. Median values and first and third quartiles of the sub-sample of 120 nights and the correspondent sample reconstructed by the model are very well correlated also with the bold full line representing the sample of all the nights of 2013 and 2017. In particular PWV is $\leq$ 1~mm for 13\% of time on 2013 and for 12\% of time on 2017. Very similar values to what has been observed on the whole years 2013 and 2017 (Fig.\ref{fig:parcumdists} - bold line). \\

As we anticipated in Section \ref{sec:obs}, even if the first two samples appear representative of the global distribution, to better investigate the most challenging region of PWV $\leq$ 1~mm, we analyzed the third sample obtained selecting all the nights in 2017 having at least 30\% of the night in which the PWV $\leq$ 1mm as described in Section \ref{sec:obs}. We identified 35 nights (Table \ref{sample201735n}).  On the third sample of 35 nights we have $\sim$73\% of measurements below or equal to 1~mm, with almost all of the measurements below or equal to 2~mm.

In Fig.\ref{fig:lowscatter} and Table \ref{tab:lowstat} we report the statistical indicators computed on the 35 nights sample. Looking at the results after correction, we obtain the excellent values of RMSE=0.27~mm and a $\sigma$=0.25~mm. This value is comparable to the level of accuracy obtained by different instrumentation (VISIR, CRIRES, UVES, X-SHOOTER) that is of the order of 0.1-0.2~mm \citep{kerber2010} - Table 4 and \citep{kerber2014} - Table 1. We can conclude therefore that the level of the model performances is very satisfactory. If we put all together the 35 nights of 2017 with the sample of 120 nights of 2017\footnote{Note that the total sample is done by 135 nights (and not 155) because part of the 35 nights are already included in the 120 nights. The two sub-samples overlap in part.} and we calculate the contigency table (see Table \ref{tab:cont2017_135}, we obtain a POD$_{1}$=84.3\% that is very similar to the POD$_{1}$=86.2\% (Table \ref{tab:cont2017}). This is consistent with the fact that we had proven that the sample of 120 nights of 2017 is very representative of the typical conditions above Cerro Paranal. \\

Finally, Table \ref{tab:pod1stats} shows the POD for PWV $\leq$ 2~mm, 1~mm and 0.5~mm obtained with the sample of 135 nights.  In the last column, we consider the POD calculated assuming an uncertainly of 0.2~mm. As we will see in Section \ref{sec:spec}, this corresponds to the typical uncertainty of instruments measuring the PWV. Besides to LHATPRO, there are also other instruments: VISIR, XSHOOTER, UVES, CRIRES). This means that we consider the Meso-NH forecast as a hit if it falls within 0.2~mm of the corresponding measurement. We conclude that the Meso-NH model has a probability of correct detection for PWV $\leq$ 2~mm of 97\%, a probability of correct detection for PWV $\leq$ 1~mm of 93\% and a probability of correct detection for PWV $\leq$ 0.5~mm of 79\%.

\begin{table}
 \begin{center}
 \caption{Cerro Paranal - In each column are the statistical indicators computed both on the raw model output and by applying correction in Eq. \ref{eq:regression}. We did no special optimization for the low PWV values. The statistics are computed over the 35 nights 2017 sample obtained by selecting the nights with episodes of low PWV$\leq$ 1~mm. The indicators are computed by filtering the sample selecting LHATPRO measurements with PWV$\leq$ 1~mm.} 
% \resizebox{\columnwidth}{!}{
 \begin{tabular}{c|ccc}
 \hline
 \multicolumn{1}{c}{ } & RMSE (mm) & BIAS (mm) & $\sigma$ (mm) \\
 \multicolumn{1}{c}{ } & (raw/corr) & (raw/corr) & (raw/corr) \\
 \hline
   & & \\
   PWV$\leq 1$~mm &  0.46/0.27 & 0.37/0.09 & 0.27/0.25 \\
   & & \\
 \hline
 \end{tabular}
% }
 \label{tab:lowstat}
 \end{center}
 \end{table}
 
  \begin{table}
 \begin{center}
 \caption{3$\times$3 contingency table - 2017 135 nights sample with applied correction in Eq. \ref{eq:regression}} 
 \resizebox{\columnwidth}{!}{
 \begin{tabular}{cc|ccc}
 \hline
 \multicolumn{2}{c}{PWV (mm)} & \multicolumn{3}{c}{\bf LHATPRO}\\
 \multicolumn{2}{c}{ } & PWV$\leq$      1 &      1 $<$PWV$\leq$      5 &  PWV$>$      5 \\
 \hline
  \multirow{7}{*}{\rotatebox{90}{\bf MODEL}} & & &\\
   & PWV$\leq$         1 &           581      &           217      &             0     \\
   & & & & \\
   &          1 $<$PWV$\leq$      5 &           108      &          1832      &            95     \\
   & & & & \\
   & PWV$>$         5 &             0      &            92      &           698     \\
   & & & & \\
 \hline
 \\
 \multicolumn{5}{l}{Sample size =       3623; PC=     85.9\%; EBD=      0.0\%} \\
 \multicolumn{5}{l}{POD$_1$=     84.3\%; POD$_2$=     85.6\%; POD$_3$=     88.0\%} \\
 \end{tabular}
 }
 \label{tab:cont2017_135}
 \end{center}
 \end{table}

\begin{table}
 \begin{center}
 \caption{Cerro Paranal - POD values computed on the 135 nights 2017 sample for PWV $\leq$ 2~mm, 1~mm and 0.5~mm. Meso-NH values are corrected by Eq. \ref{eq:regression}. In the last column we consider a 0.2~mm tolerance on the selected intervals in order to take into account the uncertainty on measurements. This means that we consider the Meso-NH forecast as a hit if it falls within 0.2~mm of the corresponding measurements. }
 %\resizebox{\columnwidth}{!}{
 \begin{tabular}{c|cc}
 \hline
 \multicolumn{1}{c}{PWV range} & POD& POD ($\pm$0.2~mm) \\
 \hline
   & \\
   \hspace{1cm} PWV$\leq 0.5$~mm &  64.8\% & 79.2\%\\
   \hspace{1cm} PWV$\leq 1$~mm  &  84.3\%& 93.5\%\\
   \hspace{1cm} PWV$\leq 2$~mm &  95.9\%& 97.3\%\\
   & \\
 \hline
 \end{tabular}
 %}
 \label{tab:pod1stats}
 \end{center}
 \end{table}
 
%\begin{table}
 %\begin{center}
% \caption{Cerro Paranal - POD values computed on the 135 nights 2017 sample for PWV weaker than 2~mm, 1~mm and 0.5~mm. Meso-NH values are corrected by Eq. \ref{eq:regression} and we consider a 0.2~mm tolerance on the selected intervals in order to take into account the model error as shown in Table \ref{tab:lowstat}. This means that we consider the Meso-NH forecast as a hit if it falls within 0.2~mm of the corresponding LHATPRO measurement. } 
 %\resizebox{\columnwidth}{!}{
% \begin{tabular}{c|c}
% \hline
% \multicolumn{1}{c}{PWV range} & POD \\
% \hline
%   & \\
%   \hspace{1cm} PWV$\leq 0.5$~mm \hspace{1cm} & \hspace{1cm} 79.24\% \hspace{1cm} \\
%   \hspace{1cm} PWV$\leq 1$~mm \hspace{1cm} & \hspace{1cm} 93.47\% \hspace{1cm} \\
%   \hspace{1cm} PWV$\leq 2$~mm \hspace{1cm} & \hspace{1cm} 97.35\% \hspace{1cm} \\
%   & \\
% \hline
% \end{tabular}
 %}
% \label{tab:pod1stats_corrected}
% \end{center}
% \end{table}
%%%%%%%%%%%%%%%%%%%%%%%%%%%%%%%%%%%%%%%%%%%%%
\subsection{Meso-NH model vs. ECMWF}
\label{sec:compar}

In this section we compare the performances of the Meso-NH model with the General Circulation Model of the European Center for Medium Range Weather Forecasts (ECMWF). We extract ECMWF historical data from the MARS catalogue\footnote{\url{https://software.ecmwf.int/wiki/display/WEBAPI/MARS+service}}. We consider the sequence of forecasts calculated at 00:00 and 12:00 hours UT with a time sampling of one hour (i.e. a data point each hour). We pick-up the ECWMF forecasts as if we are in an operational configuration i.e. we consider that data are made available by ECMWF approximately 6 hours after the calculation time. For each date J, extracted data corresponding to hours 6:00-17:00 UT are forecasts calculated at 00:00 UT, while data corresponding to 18:00-05:00 UT (of day after) are calculated at 12:00 UT of the same day J, following the 6 hours delay of the ECMWF data delivery. This corresponds to the forecast system delivered by ECMWF to ESO in the 2017 period.\\

\begin{figure*}
\centering
%\begin{adjustbox}{max width=\textwidth}
\includegraphics[width=0.8\textwidth]{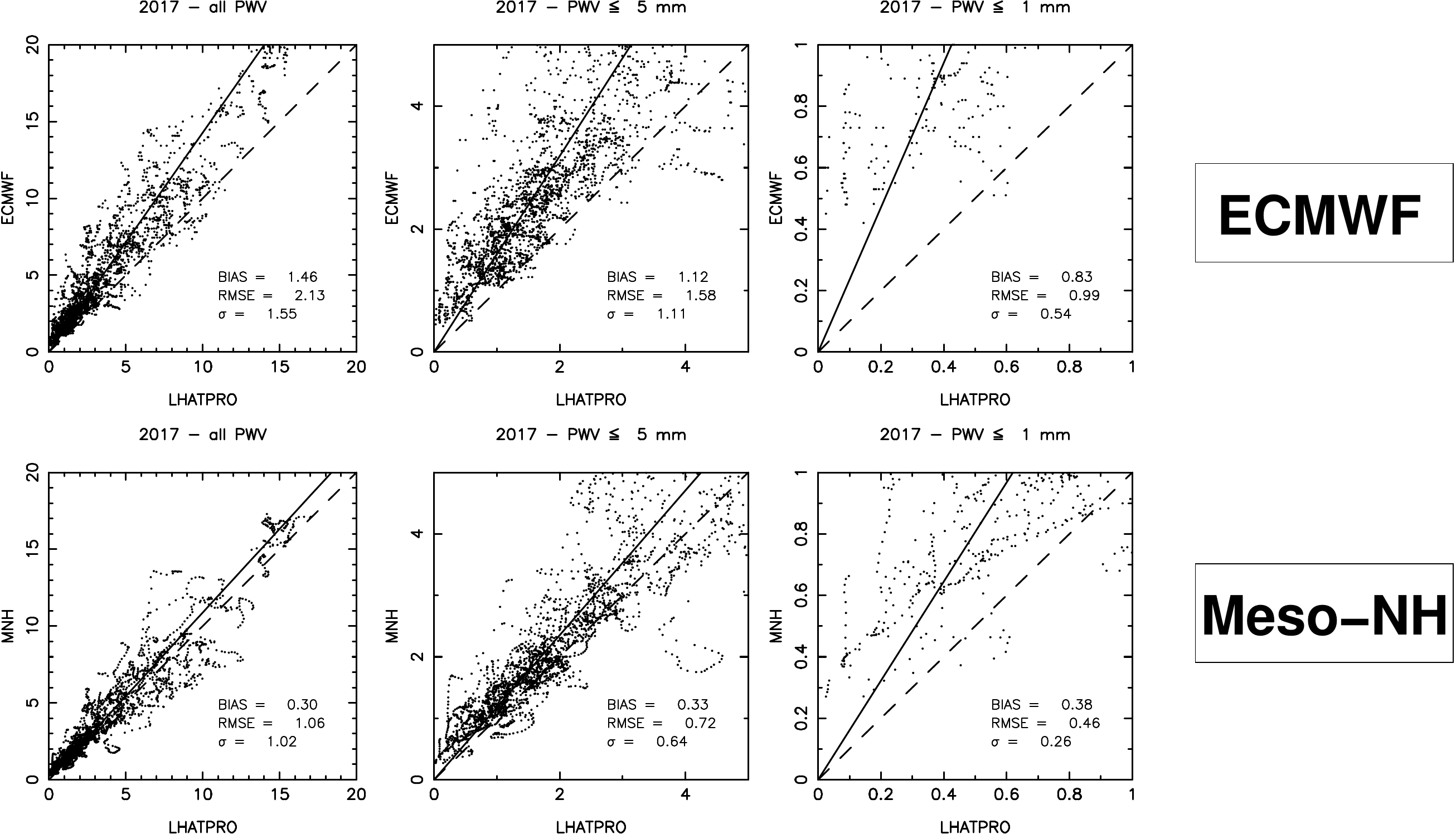}
%\end{adjustbox}
\caption{Cerro Paranal - Scatterplot computed on the 135 nights 2017 sample comparing LHATPRO measurements either with ECMWF GCM forecasts (\textbf{top row}) and Meso-NH forecasts (\textbf{bottom row}). The plots are relative to the whole sample (first column), to a sample filtered by selecting LHATPRO measurements with PWV$\leq$ 5~mm (second column) and PWV$\leq$ 1~mm (third column). The dashed line corresponds to the 45$^\circ$ that should represent a perfect match between model and measurements. The straight line corresponds to the regression line computed on the data points.}
\label{fig:ecmwfscatter2}
\end{figure*}

In the comparison between ECMWF and Meso-NH forecasts, data have been resampled every 20 minutes in order to compute a statistical analysis in identical conditions to the previous studies. we performed the comparaison on the largest possible sample i.e. the 135 nights sample (see Section \ref{sec:respwv}). In Fig. \ref{fig:ecmwfscatter2} we show the scatterplots of the PWV in the three regimes (all values, PWV $\leq$ 5~mm, PWV $\leq$ 1~mm) between ECMWF (top row) and Meso-NH (bottom line). 
In Table \ref{tab:ecmwfstat2} are reported the corresponding statistical operators. The comparison ECMWF vs. Meso-NH is done on the raw measurements without the post-processing correction (see Eq.\ref{eq:regression}) because the important information, in this context, is the difference between the two models in the same conditions. The goal of this exercise is to establish if there is a gain or not in using Meso-NH. All the statistical operators (BIAS, RMSE and $\sigma$) are visibly larger for ECMW than Meso-NH. The ECMWF BIAS is a factor between 2 and 4 larger than the Meso-NH one, while ECMWF RMSE and $\sigma$ are a factor 2 larger.\\

The most critical parameters are, however, the RMSE and even more $\sigma$ that represents the pure statistical error. In all the regimes, RMSE and $\sigma$ are visibly larger for the ECMWF case than for the Meso-NH one. In the most challenging case i.e. where the PWV $\leq$ 1~mm, we have a $\sigma$=0.54~mm (ECMWF) and $\sigma$=0.26~mm (Meso-NH) i.e. a not negligible factor 2 in gain. For RMSE, the ECMWF case (0.99) is a factor 2.15 larger than Meso-NH (0.46).\\

Such result proves that Meso-NH can provide a not negligible improvement with respect to ECMWF in terms of scheduling of VLT instruments depending on PWV estimates.\\

\begin{table}
 \begin{center}
 \caption{Cerro Paranal - In each column are the statistical indicators computed by comparing LHATPRO measurements on the 135 nights 2017 sample either with the ECMWF GCM forecasts and the Meso-NH forecasts. The indicators are computed over the whole sample (All PWV), on a sample filtered by selecting LHATPRO measurements with PWV$\leq$ 5~mm and finally on a sample filtered with PWV$\leq$ 1~mm.} 
% \resizebox{\columnwidth}{!}{
 \begin{tabular}{c|ccc}
 \hline
 \multicolumn{1}{c}{PWV (mm)} & RMSE (mm) & BIAS (mm) & $\sigma$ (mm) \\
 \multicolumn{1}{c}{ } & {\tiny (ECMWF/MNH)} & {\tiny (ECMWF/MNH)} & {\tiny (ECMWF/MNH)} \\
 \hline
   & & \\
   All PWV & 2.01/1.06 & 1.35/0.30 & 1.49/1.02 \\
   & & \\
   PWV$\leq$5 &  1.45/0.72 & 1.04/0.33 & 1.02/0.64 \\
   & & \\
   PWV$\leq$1 &  0.99/0.46 & 0.83/0.38 & 0.54/0.26 \\
   & & \\
 \hline
 \end{tabular}
% }
 \label{tab:ecmwfstat2}
 \end{center}
\end{table}

%%%%%%%%%%%%%%%%%%%%%%%%%%%%%%%%%%%%%%%%%%%%%
\subsection{Model performances in reconstructing the absolute humidity vertical profiles AH}
\label{sec:resah}

\begin{figure}
\centering
%\begin{adjustbox}{max width=\textwidth}
\includegraphics[width=0.7\columnwidth,angle=270,origin=c]{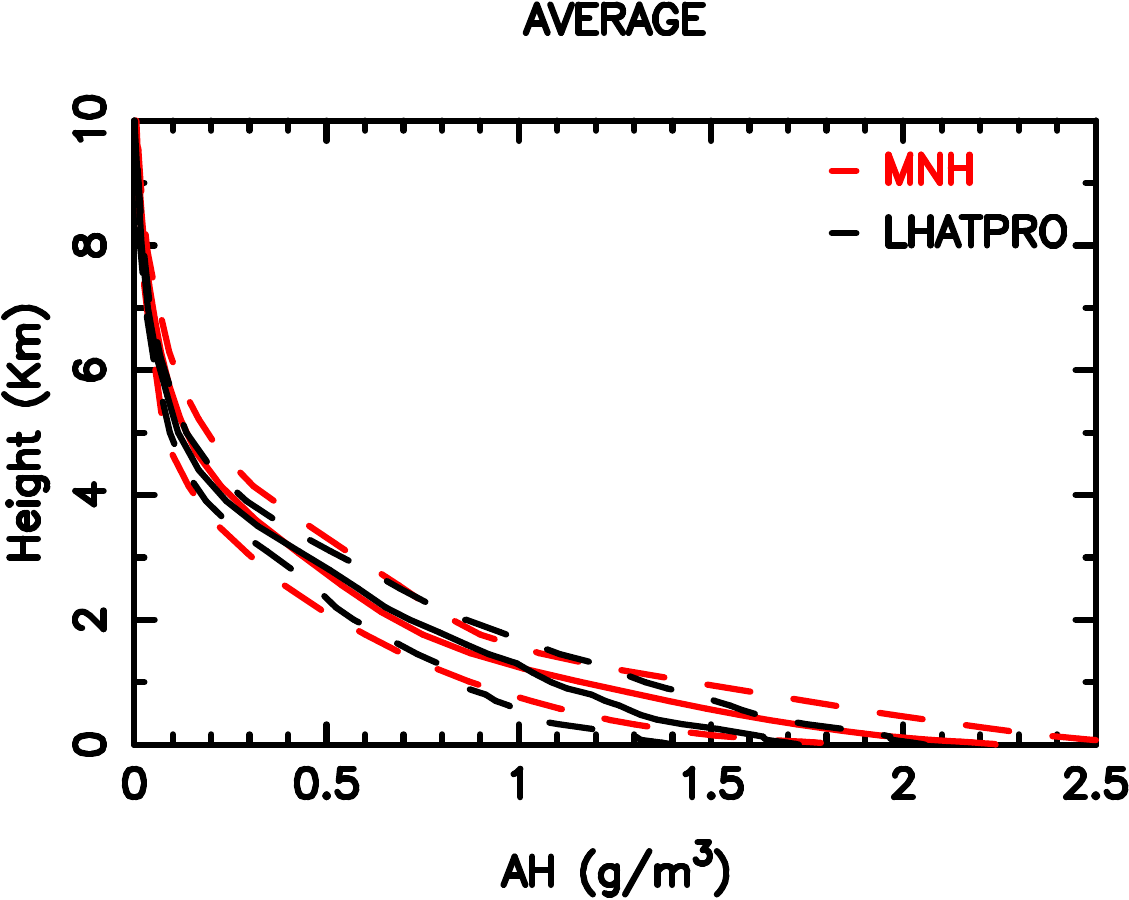}
%\end{adjustbox}
\caption{Cerro Paranal - Average vertical AH profiles computed over the 120 nights 2013 sample by the Meso-NH model (red) and by LHATPRO (black line). Dash-lines indicate the standard deviation for the model (red) and LHATPRO (black).}
\label{fig:ahprof}
\end{figure}

While not being crucial for astronomical observations, we study here the absolute humidity (AH) vertical profiles as measured by LHATPRO and reconstructed by Meso-NH because this can provide us informations on where there is space for improving model performances. We used the sample 120 nights of 2013 because it is the unique one for which we have the vertical stratification of measurements on the about 20~km above the ground. We compared therefore the AH profiles measured by LHATPRO with those reconstructed by the model  obtained with Eq.\ref{eq2}. In figure \ref{fig:ahprof} is shown the average AH profile obtained by both LHATPRO and Meso-NH, with the variability over the whole sample expressed as standard deviation and represented by the dashed lines. Above 10~km a.g.l, the water content in the atmosphere is almost negligible and does not contribute in a sensible way to the total PWV value. The agreement is meaningful because it shows that the model is correctly reproducing the highly variable distribution of water vapour across the atmosphere. Only below 1~km the model tends to overestimate, even if within the standard deviation margins. This finding is consistent with what observed in previous sections in the statistical analysis of PWV in which it was evident a residual BIAS (very small in reality) that we have corrected in a post-processing phase and can be introduced in the operational configuration. This result tells us that there is still some space for further improvement of the model in the very low part of the atmosphere where it is dominant the soil-atmosphere interaction. This might be a topic for future investigations. However we have shown in the previous sections that at even at present we can overcome this shortcoming efficiently through a statistical approach in a post-processing phase. This element has, however, has basically no impact on the forecasts model performances because we showed that can be statistically corrected in post-processing phase in an efficient way.\\

%The agreement is excellent, meaning that the model is correctly reproducing the humidity behaviour in the whole atmosphere, however we see that below 1~km a.g.l. the model tends to overestimate slightly the water content, even if within the standard deviation margins. This finding is consistent to what observed in previous sections in the statistical analysis of PWV in which it was evident a residual BIAS (very small in reality) that we have corrected in a post-processing phase and can be introduced in the operational configuration. This result tells us that there is still some space for further improvement of the model in the very low part of the atmosphere where it is more evident the soil-atmosphere interaction. {\bf This might be topic for future investigations.} However we have shown in the previous sections that at even at present we can overcome this shortcoming efficiently through a statistical approach in a post-processing phase. \\

%This element has, however, has basically no impacts on the forecasts model performances because we showed that can be statistically corrected in post-processing phase in an efficient way.\\

%%%%%%%%%%%%%%%%%%%%%%%%%%%%%%%%%%%%%%%%%%%%%
\subsection{Forecast of an episode of extremely low PWV}
\label{sec:spec}

\begin{figure*}
\centering
%\begin{adjustbox}{max width=\textwidth}
\includegraphics[width=1.0\textwidth]{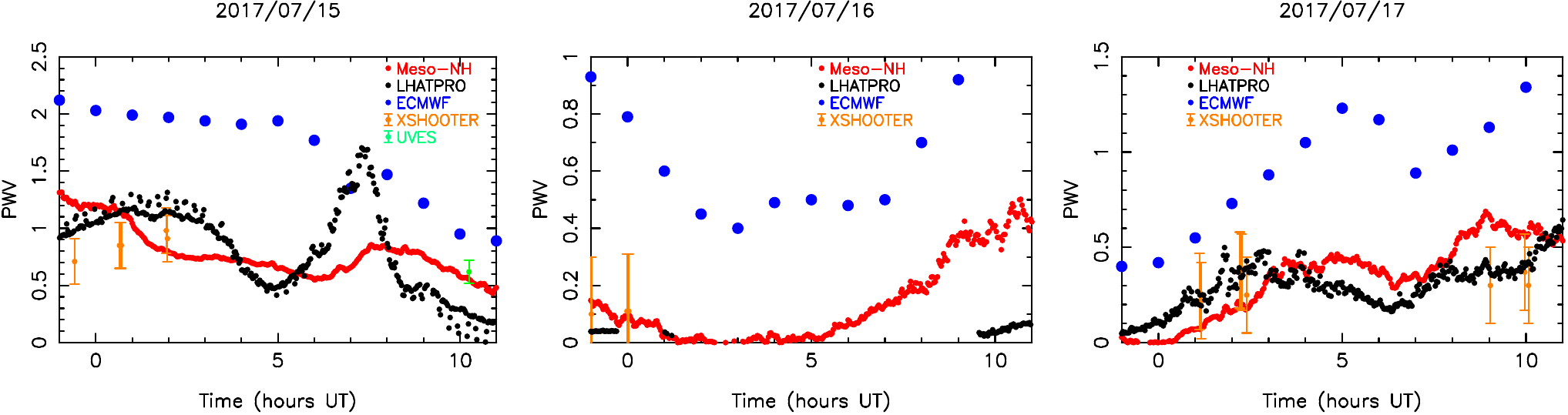}
%\end{adjustbox}
\caption{Time evolution of PWV values over three nights centered arount the extremely low PWV event of 2017/07/17. Black dots correspond to LHATPRO measurements, red dots correspond to the Meso-NH forecast (2-minutes sampling time), the blue dots corresponds to the ECMWF forecast (1-hour sampling time), the orange dots corresponds to X-SHOOTER measurements and green dots corresponds to UVES measurements. According to Table \ref{tab:lowstat}, we estimate the Meso-NH model error to be around 0.25 mm.}
\label{fig:case}
\end{figure*}

We show here results obtained in relation to a single extremely low PWV event registered at Cerro Paranal in 2017. PWV show extremely low values in one night in a period of three consecutive nights [2017/07/15-2017/07/17] UT\footnote{Our convention on dates is that yyyy/mm/dd UT indicates the start of the local night.} in which the PWV was lower than 1 mm reaching values smaller than 0.025~mm\footnote{when the measurement is not visible in the figure it means that PWV is basically equal to zero as it will be explained later on} on 2017/07/16. Fig.\ref{fig:case} show the time evolution of the PWV during the three nights. Predictions of the Meso-NH model are shown together with measurements from the radiometer LHATPRO and from other VLT instruments. During these three nights we have the availability of PWV measurements from UVES and X-SHOOTER instruments. X-SHOOTER is a multi-wavelength medium resolution spectrograph and operates on wavelength between 300 nm and 2500 nm \citep{vernet2011}. We refer to \citet{kerber2014} for more details on the instruments X-SHOOTER and UVES.

%It is located on UT2 facility at VLT and its measurements are influenced by the water absorption in a progressive way moving further into the infrared domain. PWV estimated are retrieved by observing telluric standard stars at airmass similar to the scientific object and within 2h time from the scientific observation. A reference spectrum of telluric absorption within 2h of each scientific spectrum is measured in the 200-2500 nm range. A pipeline measures then an equivalent width of telluric absorption lines in the range 717-720 nm and deduces PWV through a curve of growth. A similar method is used for UVES \citep{kerber2014}.\\

In Fig.\ref{fig:case} we have therefore a sampling of 5~s for LHATPRO and 120~s for Meso-NH model. We considered an error bar of 0.20~mm for X-SHOOTER and 0.1 mm for UVES as reported by \citet{kerber2014}. For this special case we run longer simulations that cover a 18 hours time interval between 18 UT to 12 UT of the day after. We plotted all the available measurements from LHATPRO (black dots),  X-SHOOTER (orange dots) and UVES (green dots) over the selected nights. The model shows a very good correlation with measurements done with LHATPRO and other VLT instruments. In almost all cases the discrepancy between the model and the measurements is within the error bar of the VLT instruments. In this case the PWV measurements are in the lowest range of values (PWV$\leq$1 mm), so according to Table \ref{tab:lowstat}, we estimate the model error to be around 0.25 mm.
ECMWF forecasts (blue dots), extracted with the same procedure explained in section \ref{sec:compar}, are unable to represent the extremely low PWV values of this episode with the same accuracy of the mesoscale Meso-NH model, and this behaviour is particularly evident in the 2017/07/16 night. During this night the number of LHATPRO PWV measurements that appear on the figure is small. The apparent missing measurements correspon in reality to a PWV equal to zero because PWV retrieval from brightness temperature fails if the PWV value is too low. This means that where we do not see the black points (LHATPRO PWV measurements) we can consider PWV equal to zero. LHATPRO measurements are consistent with both X-SHOOTER and Meso-NH up to an exceptional degree. From this simple test we do expect that the Meso-NH forecast can be indeed able to predict in advance such rare events with excellent levels of accuracy and support infrared observations at VLT. Also Meso-NH shows definitely better performances with respect to the ECMWF predictions. We also highlight the fact that, besides the evident better performances of Meso-NH with respect to ECMWF, the temporal sampling of the former (2~min) is definitely better than the latter (1h).\\

%%%%%%%%%%%%%%%%%%%%%%%%%%%%%%%%%%%%%%%%%%%%%
\section{Model validation on Mt.Graham}
\label{sec:mtg}

In the case of LBT Observatory (Mount Graham), we do not have an instrument similar to LHATPRO deployed on the site. We cannot therefore have access to accurate measurements of PWV and the vertical AH stratification.  However, the excellent results obtained in our analysis performed over Cerro Paranal make us confident that the model can perform reasonably well also on Mount Graham, since they are similar sites (mountain tops) and no specific model calibration as that required for the optical turbulence \citep{masciadri2017} is done before running the simulations. We assume that the coefficient of correction calculated a posteriori (Eq.\ref{eq:regression}) are valid not only for Cerro Paranal but also for other sites. The PWV is indeed weakly dependent on the topography. The fact that, as we will see, model outputs are consistent with measurements at Mt. Graham from a climatologic point of view, indicates that this assumption is reasonable. It is in any case obvious that, the presence of measurements in situ, might permit a more sophisticated validation at Mt. Graham similar to that done for Cerro Paranal. To provide at least a climatological verification, we compare here the climatologic statistical values of operators characterizing the PWV as reconstructed by the Meso-NH model from the ALTA Center project and as measured from GOES satellites measurements by \citet{carrasco2017} on a total of 58 months between 1993 and 1999. ALTA Center is also charcaterized by a three domains structure as is the case described here for Cerro Paranal having the same horizontal resolution of 10~km, 2.5~km and 0.5~km\footnote{A fourth domain with higher horizontal resolution is included but it is used only for the wind speed.}. 

In \citet{carrasco2017} the PWV values are reconstructed from the brigthness temperature at 6.7 $\mu$m, measured by satellites, through a semi-empirical method developed by \citet{soden1993}. 
%This technique requires clear-sky conditions to be performed, so it may leads to an underestimation of PWV climatology due to the fact that in presence of clouds measurement is impossible. 
In Section 3.3.1 of the Carrasco paper, the technique was validated on Mount Graham against radiometer measurements provided by the Submillimeter Telescope Observatory (SMTO). The comparison was done by considering only radiometer values less than 7~mm (which ensures the validity of the comparison) and the found agreement was good.\\
Thanks to the ALTA system already running on the telescope site since last year, we have the availability of 283 simulated nights from 2016/09/21 (date in which the PWV forecast was initially implemented in ALTA) to 2017/06/30 (UT dates). This is basically one solar year if we exclude the July-August summer shutdown period of the telescope. 

\begin{table}
 \begin{center}
 \caption{Comparison between the values ($1^{st}$ quartile, median and $3^{rd}$ quartile) as obtained by the Meso-NH model (from ALTA Center project) on one solar year and the values as retrieved by Carrasco et al. on a sample of 58 months between 1993 and 1999. See text for more details.} 
% \resizebox{\columnwidth}{!}{
 \begin{tabular}{c|ccc}
 \hline
 \multicolumn{1}{c}{ } & $1^{st}$ quartile & Median & $3^{rd}$ quartile \\
 \multicolumn{1}{c}{ } & (mm) & (mm) & (mm) \\
 \hline
   & & \\
   ALTA & 1.9 & 2.9 & 4.1 \\
   & & \\
   Carrasco et. al & 2.0 &  2.9 & 4.3 \\
   & & \\
 \hline
 \end{tabular}
% }
 \label{tab:mtgstat}
 \end{center}
\end{table}

\begin{figure*}
\centering
%\begin{adjustbox}{max width=\textwidth}
\includegraphics[width=1.0\textwidth]{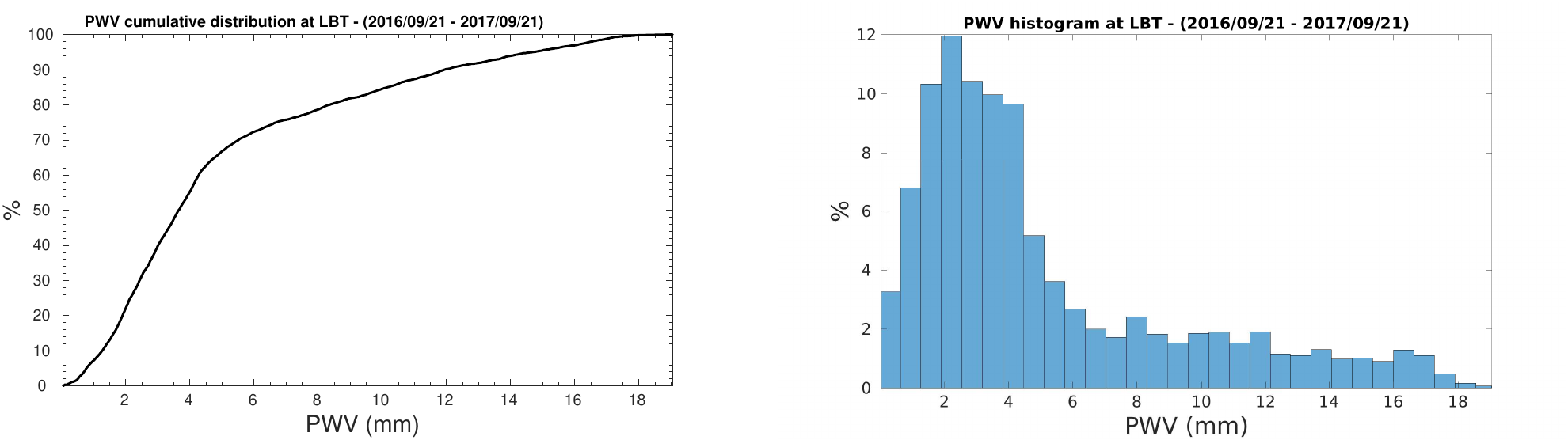}
%\end{adjustbox}
\caption{Cumulative distribution (left) and histogram (right) of PWV values on Mount Graham forecasted by Meso-NH model as provided by the ALTA Center. Data refers to 283 nights between 2016/09/21 and 2017/09/21.}
\label{fig:mtg}
\end{figure*}

In Fig.\ref{fig:mtg} is shown the distribution of PWV values produced by the ALTA system over the selected one-year period, corrected by Eq. \ref{eq:regression} that was obtained by the analysis on Cerro Paranal. We assume that the correction can be applied everywhere because in principle we did not adjust the model parameters with a site-dependent calibration. From the statistics computed over the simulated period, if we consider only values of PWV $\leq$ 7~mm as indicated by \citet{carrasco2017} to respect the validation conditions, we obtain a $1^{st}$ quartile, median and $3^{rd}$ quartile of the distribution reported in Table \ref{tab:mtgstat}. The agreement with Carrasco's values is very good and confirms the validity of ALTA forecasts on Mount Graham. We note that, considering the nature of the GOES satellites estimates that we are treating, we can not perform more accurate analyses. It has been already noted that GOES satellites is preferably used for statistical analysis since the scatter on individual nights is very high \citep{kerber2010}. In part because we can not perform a comparison on the same specific nights but we can only perform a comparison in climatological sense as described here. While this kind of validation is not as detailed as in the Paranal case and it can mainly achieve a climatologic estimate, the results we obtained in Table \ref{tab:mtgstat} indicate that Meso-NH provides very consistent estimate for the PWV above Mt.Graham. 

%%%%%%%%%%%%%%%%%%%%%%%%%%%%%%%%%%%%%%%%%%%%%
\section{Conclusions}
\label{concl}

In this paper we validated the Meso-NH model in forecasting the PWV above Cerro Paranal, site of the VLT and Mt.Graham, site of the LBT. A detailed analysis has been performed at Cerro Paranal with two rich statistical sample of nights (120 nights) in two different years (2013 and 2017) showing excellent performances with the best on 2017 that corresponds to the best initialization data. To better investigate the portion of PWV values below 1~mm we also considered a slighter richer sample on 2017 (135 nights) including all the night of 2017 having a PWV $\leq$ 1~mm for at least 30\% of the night.
 
The statistical operators (BIAS, RMSE and $\sigma$) applied on the sample of all PWV values, on the PWV $\leq$ 5~mm and on the PWV $\leq$ 1~mm are all consistently extremely good with the RMSE and $\sigma$ that decrease when the extension of the interval of analysis (PWV $\leq$ X~mm) is limited to smaller X values reaching the smallest values of RMSE = 0.27~mm and $\sigma$ = 0.25~mm in the most challenging region of PWV $\leq$ 1~mm. \\
We demonstrated that the Meso-NH model provides a significant improvement with respect to the ECMWF forecast (roughly a factor 2) for the statistical operators (RMSE and $sigma$) passing, respectively from 1.58~mm to 0.72~mm (for PWV $\leq$ 5~mm) and from 0.99~mm to 0.46~mm (for PWV $\leq$ 1~mm) for RMSE and from 1.11~mm to 0.54~mm (for PWV $\leq$ 5~mm)  and from 0.54~mm to 0.26~mm (for PWV $\leq$ 1~mm) for $\sigma$. 

We also calculated 3$\times$3 contingency tables with the correspondent percentage of corrected prediction (PC) and probability of detection in each individual sectors: PWV $\leq$ 1~mm, 1~mm $<$ PWV $\leq$ 5~mm and PWV $>$ 5~mm. We investigated finally the probability of detection for the most challenging sectors: $\leq$ 2~mm, $\leq$ 1~mm and $\leq$ 0.5~mm.
The probability to detect the PWV $\leq$ 2~mm is 97\%, the probability to detect the PWV $\leq$ 1~mm is 93\% and the probability to detect the PWV $\leq$ 0.5~mm is 79\%. 

Besides, we also studied, in detail way, a case of extremely low PWV observed at the VLT lasted three days (with values smaller than 0.025~mm on one night). We showed that Meso-NH can reconstruct the trend and the values observed by LHATPRO and by further two instruments (X-SHOOTER and UVES) in this period in a much more precise way than ECWMF forecasts. The discrepancy between the model and the observations is almost comparable to the dispersion among the instruments and in many cases also with the declared accuracy of the instruments. 

By studying the vertical stratification of the absolute humidity as measured by LHATPRO and reconstructed by the model we can conclude that the Meso-NH model provides a vertical median profile very well correlated to measurements. Only very close to the surface, it is evident a slight overestimation of the model that is, anyway, within the standard deviation. This tells us that in this region of the atmosphere we might have some small margin of improvements for the model. 

Once the model has been validated in an extremely detailed way above Cerro Paranal (VLT), we compared measurements coming from GOES satellites \citep{carrasco2017} and simulations performed by Meso-NH. This comparison has been done in climatological terms by comparing model estimates on one solar year with satellites measurements related to 58 months in the period 1993-1999. Comparison provided an excellent agreement on the median value (2.9~mm for both measurements and Meso-NH) and first (1.9~mm and 2.0~mm) and third (4.1~mm and 4.3~mm) quartiles of the distribution for PWV $\leq$ 7~mm that, as proved by \citet{carrasco2017}, is the range of validity of the method.

The model Meso-NH appears therefore an extremely efficient tool to forecast the PWV above Cerro Paranal (VLT), Mt.Graham (LBT) and similar astronomical sites. The investigation of model performances on longer time scales (larger than 15~h) is not a priority at the moment. On the other side we are working on techniques/methods to improve the model behaviour on short time scale (a few hours), a goal that is very critical for the flexible scheduling.
This study represents an important step towards the set-up of a system for the operational forecast of various atmospheric parameters, among which the PWV, that is in progress by a few of us at INAF and it is conceived for the VLT.

%%%%%%%%%%%%%%%%%%%%%%%%%%%%%%%%%%%%%%%%%%%%%
\section*{Acknowledgements}

ALTA Center project is funded by the Large Binocular Telescope Corporation. The authors thanks Christian Veillet, Director of the Large Binocular Telescope, for his continued and valuable support given to this research activity. Authors also thanks the LBTO staff for their technical support and collaboration. The author thanks the MOSE ESO Board and the Paranal Science Operation Team for their constant support. Part of the numerical simulations have been run on the HPCF cluster of the European Center for Medium Range Weather Forecasts (ECMWF) using resources from the Project SPITFOT.

%%%%%%%%%%%%%%%%%%%%%%%%%%%%%%%%%%%%%%%%%%%%%%%%%%



%\bibliographystyle{mnras}

\begin{thebibliography}{99}

\bibitem[\protect\citeauthoryear{Bougeault et al.}{1989}]{Bougeault89}
Bougeault, P. \& Lacarr\`ere, P., 1989, Mon. Weather. Rev., 117, 1872-1890

\bibitem[\protect\citeauthoryear{Carpenter}{1982}]{somm}
Carpenter, K. M., 1982, Quart. J. Roy. Meteor. Soc., 110, 717-719

\bibitem[\protect\citeauthoryear{Carrasco et al.}{2017}]{carrasco2017}
Carrasco, E., Avila, R., Erasmus, A., Djorgovski, S.G., Walker, A.R., Blum, R., 2017, PASP, 129, 973

\bibitem[\protect\citeauthoryear{Cuxart et al.}{2000}]{Cuxart00}
Cuxart, J., Bougeault, P., Redelsperger, J.-L., 2000, Q. J. R. Meteorol. Soc., 126, 1-30

\bibitem[\protect\citeauthoryear{Eisenhauer et al.}{2008}]{eisenhauer2008}
Eisenhauer, F., Perrin, G., Rabien, S.,  Eckart, A., Lena, P., Genzel, R., Abuter, R., Paumard, T., Brandner, W., 2008, ESO Astrophysics Symposia, 41, 431

\bibitem[\protect\citeauthoryear{Gal-Chen et al.}{1975}]{chen1975}
Gal-Chen, T., and R. C. J. Sommerville, 1975, J. Comput. Phys., 17, 209-228

%\bibitem[\protect\citeauthoryear{Giordano et al.}{2013}]{giordano2013}
%Giordano, C., Vernin, J., Vázquez Ramió, H., Muñoz-Tuñón, C., Varela, A. M., Trinquet, H., 2013, MNRAS, 430, 3102

\bibitem[\protect\citeauthoryear{Giordano et al.}{2013}]{giordano2013}
Giordano, C., Vernin, J., Vazquez Ramio, H., Munoz-Tunon, C., Varela, A. M., Trinquet, H., 2013, MNRAS, 430, 3102

\bibitem[\protect\citeauthoryear{Giovanelli et al.}{2001}]{giovanelli}
Giovanelli, R., Darling, J., Henderson, C., Hoffman, W., Barry, D., Cordes, J., Eikenberry, S., Gull, G., Keller, L., Smith, J. D., Stacey, G., 2001, PASP, 113, 803

\bibitem[\protect\citeauthoryear{Henault et al.}{2003}]{henault2003}
Henault, F., Bacon, R., Bonneville, C., Boudon, D., Davis, R.L., Ferruit, P., Gilmore, G.F., Le Fevre, O., Lemonnier, J.P., Lilly, S., Morris, S.L., Prieto, E., Steinmetz, M., de Zeeuw, P.T., 2003, Proc. SPIE, 4841, 1096

\bibitem[\protect\citeauthoryear{Hinz et al.}{2014}]{hinz}
Hinz, P., Bailey, V. P., Defrere, D., Downey, E., Esposito, S., Hill, J., Hoffmann, W. F., Leisenring, J.,
Montoya, M., McMahon, T., Puglisi, A., Skemer, A., Skrutskie, M., Vaitheeswaran, V., and Vaz, A., 2012, 91460T-91460T-10

\bibitem[\protect\citeauthoryear{Hogan et al.}{2016}]{hogan}
Hogan, R. J. and Bozzo, A., 2016, ECMWF Technical Memorandum, Tech. Rep. 787

\bibitem[\protect\citeauthoryear{Jarvis et al.}{2008}]{srtm}
Jarvis, A., H.I. Reuter, A. Nelson, E. Guevara, 2008, Hole-filled SRTM for the globe Version 4, available from the CGIAR-CSI SRTM 90m Database

\bibitem[\protect\citeauthoryear{Kaeufl et al.}{2004}]{kaeufl2004}
Kaeufl, H.U., Ballester, P., Biereichel, P., Delabre, B., Donaldson, R., et al., 2004, Proc. SPIE, 5492, 1218

\bibitem[\protect\citeauthoryear{Kerber et al.}{2010}]{kerber2010}
Kerber, F., Querel, R.R. Hanuschik, R.W., Chach\'on, A., Caneo, M., Cortes, L., Cure, M., Illanes, L., Naylor, D.A., Smette, A., Sarzin, M. Rabanus, D., Tompkins, G., 2010, Proc. SPIE, 7733, 77331M

\bibitem[\protect\citeauthoryear{Kerber et al.}{2012}]{kerber2012}
Kerber, F., Rose, T. Chacon, A., Cuevas, O., Czekala, H., Hanuschik, R., Momany, Y., Navarette, J., Querel, R. R., Smette, A., van den Ancker, M., Cur\'e, M., Naylor, D. A., 2012, Proc. SPIE, 8448, 84463N

\bibitem[\protect\citeauthoryear{Kerber et al.}{2014}]{kerber2014}
Kerber, F., Querel, R. R., Rondanelli, R., Hanuschik, R., van den Ancker, M., Cuevas, O., Smette, A., Smoker, J., Rose, T., Czekala, H., 2014, MNRAS, 439, 247

%\bibitem[\protect\citeauthoryear{Kerber}{2015}]{kerber2015}
%Kerber, F., Querel, R. R., Neureiter, B., 2015, Journal of Physics: Conference Series, 595, 012017

\bibitem[\protect\citeauthoryear{Kessler}{1969}]{kessler}
Kessler, E., 1969, Meteor. Monog., 10, n. 32, 84

\bibitem[\protect\citeauthoryear{Lac et al.}{2018}]{lac2018}
Lac, C. et al., 2018, Geosci. Model Dev., 11, 1929

\bibitem[\protect\citeauthoryear{Lafore et al.}{1998}]{lafore98}
Lafore, J.-P., Stein, J., Asencio, N., Bougeault, P., Ducrocq, V., Duron, J., Fischer, C., Hereil, P., Mascart, P., Masson, V., Pinty, J.-P., Redelsperger, J.-L.,
Richard, E., Vil\`a-Guerau de Arellano, J., 1998, Annales Geophysicae, 16, 90-109

\bibitem[\protect\citeauthoryear{Lagage et al. }{2004}]{lagage}
Lagage P. O. et al., 2004, The Messenger, 117, 12

\bibitem[\protect\citeauthoryear{Lascaux et al.}{2013}]{lascaux2013}
Lascaux, F., Masciadri, E., Fini, L., 2013, MNRAS, 436, 3147

\bibitem[\protect\citeauthoryear{Lascaux et al.}{2015}]{lascaux2015}
Lascaux, F., Masciadri, E., Fini, L., 2015, MNRAS, 449, 1664

%\bibitem[\protect\citeauthoryear{Martelloni et al.}{2018}]{martelloni2018}
%Martelloni, G., Turchi, A., Masciadri, E., SPIE 

\bibitem[\protect\citeauthoryear{Morcrette et al.}{1999}]{morcrette}
Morcrette, J. J., Barker, H. W., Cole, J. N. S., Iacono, M. J., Pincus, R., 2008, Mon. Weather Rev., 136, 4773

\bibitem[\protect\citeauthoryear{Masciadri et al.}{1999}]{masciadri1999}
Masciadri E., Vernin J., Bougeault P., 1999, A\&ASS, 137, 185

\bibitem[\protect\citeauthoryear{Masciadri et al.}{2013}]{masciadri2013}
Masciadri, E., Lascaux, F., Fini, L., 2013, MNRAS, 436, 1968

\bibitem[\protect\citeauthoryear{Masciadri et al.}{2017}]{masciadri2017}
Masciadri E., Lascaux, F., Turchi, A., Fini, L., 2017, MNRAS, 466, 520

\bibitem[\protect\citeauthoryear{Mlawer et al.}{1997}]{mlawer}
Mlawer, E. J., Taubman, S. J., Brown P. D., Iacono M. J., Clough, S. A., 1997, J. Geophys. Res., 102, 16663

\bibitem[\protect\citeauthoryear{Noilhan et al.}{1989}]{Noilhan89}
Noilhan, J. \& Planton, S., 1989, Mon. Weather. Rev., 117, 536-549

\bibitem[\protect\citeauthoryear{Otarola et al.}{2010}]{otarola}
Otarola, A., Travouillon, T., Schack, M.; Els, S., Riddle, R., Skidmore, W., Dahl, R., Naylor, D., Querel, R., 2010, PASP, 122, 470

%\bibitem[\protect\citeauthoryear{Pérez-Jordán et al.}{2018}]{perez2018}
%Pérez-Jordán, G., Castro-Almazán, Muñoz-Tuñón, C., 2018, MNRAS, 477, 5477

%\bibitem[\protect\citeauthoryear{Pozo et al.}{2016}]{pozo2016}
%Pozo, D., Marín, C.,  Illanes, L., Curé, M., Rabanus, D., 2016, MNRAS, 459, 419

\bibitem[\protect\citeauthoryear{Perez-Jordan et al.}{2018}]{perez2018}
Perez-Jordan, G., Castro-Almazan, Munoz-Tunon, C., 2018, MNRAS, 477, 5477

\bibitem[\protect\citeauthoryear{Pozo et al.}{2016}]{pozo2016}
Pozo, D., Marin, C.,  Illanes, L., Cure, M., Rabanus, D., 2016, MNRAS, 459, 419

\bibitem[\protect\citeauthoryear{Primas et al.}{2016}]{Primas}
Primas, F., Marteau, S., Tacconi-Garman, L. E., Mainieri, V., Mysore, S., Rejkuba, M., Hilker, M., Patat, F., Sterzik, M., Kaufer, A., Mieske, S., 2016, Proc. SPIE, 9910, 991002

\bibitem[\protect\citeauthoryear{Querel et al.}{2016}]{Querel2016}
Querel, R. R., Naylor, D. A., Kerber, F., 2016, PASP, 123, 222

%\bibitem[\protect\citeauthoryear{Roellig et al.}{2010}]{roellig}
%Roellig T. L., Yuen L., Sisson D., Meyer A., 2010, Proc. SPIE, 7733, 773339

\bibitem[\protect\citeauthoryear{Rose et al.}{2005}]{Rose}
Rose, T. Crewell, S. Loehnert, U., Simmer, C., 2005, Atmospheric Research, 75, 183-200

\bibitem[\protect\citeauthoryear{Saunders et al.}{2009}]{saunders}
Saunders, W., Lawrence, J. S., Storey, J. W. V., Ashley, M. C. B., Kato, S., Minnis, P., Winker, D. M., Liu, G., Kulesa, C., 2009, PASP, 121, 976

\bibitem[\protect\citeauthoryear{Sims et al.}{2012}]{sims}
Sims G. et al., 2012, PASP, 124, 74

\bibitem[\protect\citeauthoryear{Smette et al.}{2008}]{smette2008}
Smette A., Horst H., Navarrete J., in Kaufer A., Kerber F., eds, 2008, ESO Astrophysics Symposia, The 2007 ESO Calibration Workshop. Springer- Verlag, Berlin, 433

\bibitem[\protect\citeauthoryear{Soden et al.}{1993}]{soden1993}
Soden, B. J., Bretherton, F. P., 1993, J. Geophys. Res., 96, 16669–88

\bibitem[\protect\citeauthoryear{Stein et al.}{2000}]{Stein00}
Stein, J., Richard, E., Lafore, J.P., Pinty, J.P., Asencio, N., Cosma, S., 2000, Meteorol. Atmos. Phys., 72, 203-221

\bibitem[\protect\citeauthoryear{Tremblin et al.}{2012}]{tremblin}
Tremblin, P., Schneider, N., Minier, V., Durand, G. Al., Urban, J., 2012, A\&A, 548, A65

\bibitem[\protect\citeauthoryear{Turchi et al.}{2017}]{turchi2017}
Turchi, A., Masciadri, E., Fini, L., 2017, MNRAS, 466, 1925

\bibitem[\protect\citeauthoryear{Vernet et al.}{2011}]{vernet2011}
Vernet J. et al., 2011, A\&A, 536, A105 

\end{thebibliography}


%%%%%%%%%%%%%%%%%%%%%%%%%%%%%%%%%%%%%%%%%%%%%%%%%%

%%%%%%%%%%%%%%%%% APPENDICES %%%%%%%%%%%%%%%%%%%%%

\newpage
\appendix

\section{Cerro Paranal climatology}
\label{climatology}
We report here the cumulative distributions of PWV obtained in 2013 and 2017 on the full year (all available LHATPRO measurements) and the 120 nights samples used in the analysis. We also report the cumulative distributions obtained by Meso-NH simulations on the same samples, with correction in Eq. \ref{eq:regression}.

% \begin{figure*}
% \centering
% %\begin{adjustbox}{max width=\textwidth}
% \includegraphics[width=0.9\textwidth]{FIG_CUMDIST-crop.pdf}
% %\end{adjustbox}
% \caption{Cerro Paranal - Cumulative distributions of PWV values in the years 2013 (left) and 2017 (right). We report the cumulative distribution computed over the LHATPRO measuraments on the whole year (full black line), the LHATPRO measurements over the 120 nights sample selected in each year (dashed black line) and the Meso-NH forecasts on the same sample (full red line), which were used to compute the statistical indicators in section \ref{sec:respwv}. In the boxes in each figure we report the median, first quartile and third quartile values computed over the respective distributions.}
% \label{fig:parcumdists}
% \end{figure*}

\begin{figure}
\centering
%\begin{adjustbox}{max width=\textwidth}
\includegraphics[width=0.9\columnwidth]{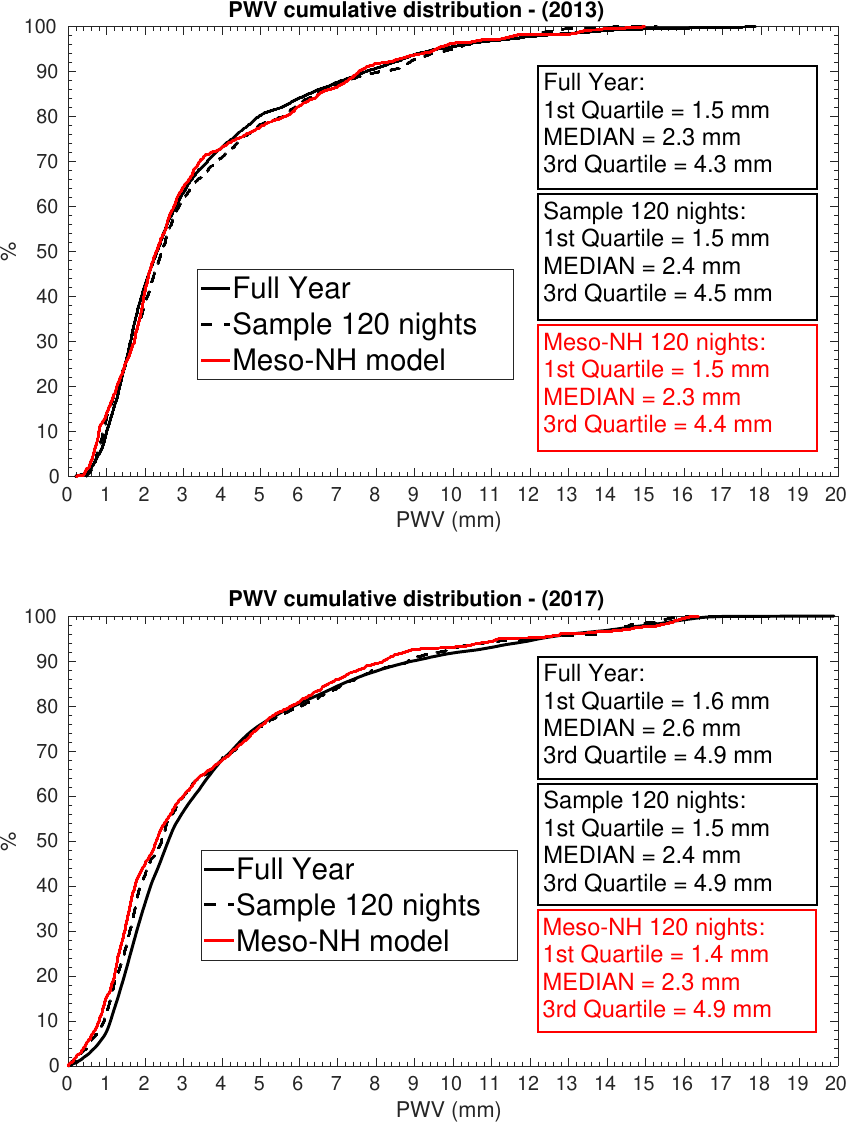}
%\end{adjustbox}
\caption{Cerro Paranal - Cumulative distributions of PWV values in the years 2013 (top) and 2017 (bottom). We report the cumulative distribution computed over the LHATPRO measuraments on the whole year (full black line), the LHATPRO measurements over the 120 nights sample selected in each year (dashed black line) and the Meso-NH forecasts on the same sample (full red line), which were used to compute the statistical indicators in section \ref{sec:respwv}. In the boxes in each figure we report the median, first quartile and third quartile values computed over the respective distributions.}
\label{fig:parcumdists}
\end{figure}

\section{Sample 35 nights 2017}
\label{sample35section}
In this annex we report the 35 nights selected in 2017, with at least 30\% of measured PWV$\leq$ 1~mm between 00:00 and 09:00 UT, which analysis is reported in figure \ref{fig:lowscatter}. The list is presented for reference in case of further studies on the selected sample of low PWV nights. The occurence of PWV$\leq$ 1~mm is more likely during the (southern) winter period, however it may happen along the whole year. Since the uneven distribution of this phenomenon through the year we selected the present sample separately from the rest.

% \begin{table*}
% \begin{center}
% \caption{List of the 35 nights in 2017 year (start date of the night in UT time, yyyy/mm/dd) with at least 30\% of measured PWV$\leq1$~mm between 00:00 and 09:00 UT.} 
% \label{tab:dates}
% %\begin{adjustbox}{max width=\columnwidth}
% \begin{tabular}{|c|c|c|c|c|c|c|c|c|}
% \hline
% 2017/04/28 & 2017/05/20 & 2017/05/26 & 2017/05/30 & 2017/07/15 & 2017/07/16 & 2017/07/17 & 2017/08/01 & 2017/08/02 \\
% 2017/08/12 & 2017/08/26 & 2017/08/27 & 2017/09/04 & 2017/09/05 & 2017/09/06 & 2017/09/07 & 2017/09/10 & 2017/09/11 \\
% 2017/09/12 & 2017/09/13 & 2017/09/14 & 2017/09/15 & 2017/09/20 & 2017/09/21 & 2017/09/25 & 2017/09/26 & 2017/10/01 \\
% 2017/10/07 & 2017/10/09 & 2017/11/17 & 2017/11/18 & 2017/11/22 & 2017/11/29 & 2017/12/07 & 2017/12/08 & \\
% \hline
% \end{tabular}
% \label{sample201735n}
% %\end{adjustbox}
% \end{center}
% \end{table*}

\begin{table}
\begin{center}
\caption{List of the 35 nights in 2017 year (start date of the night in UT time, yyyy/mm/dd) with at least 30\% of measured PWV$\leq1$~mm between 00:00 and 09:00 UT.} 
\label{tab:dates}
%\begin{adjustbox}{max width=\columnwidth}
\begin{tabular}{|c|c|c|c|}
\hline
2017/04/28 & 2017/05/20 & 2017/05/26 & 2017/05/30 \\
2017/07/15 & 2017/07/16 & 2017/07/17 & 2017/08/01 \\
2017/08/02 & 2017/08/12 & 2017/08/26 & 2017/08/27 \\
2017/09/04 & 2017/09/05 & 2017/09/06 & 2017/09/07 \\
2017/09/10 & 2017/09/11 & 2017/09/12 & 2017/09/13 \\
2017/09/14 & 2017/09/15 & 2017/09/20 & 2017/09/21 \\
2017/09/25 & 2017/09/26 & 2017/10/01 & 2017/10/07 \\
2017/10/09 & 2017/11/17 & 2017/11/18 & 2017/11/22 \\
2017/11/29 & 2017/12/07 & 2017/12/08 & \\
\hline
\end{tabular}
\label{sample201735n}
%\end{adjustbox}
\end{center}
\end{table}

%%%%%%%%%%%%%%%%%%%%%%%%%%%%%%%%%%%%%%%%%%%%%%%%%%

% Don't change these lines
\bsp	% typesetting comment
\label{lastpage}
\end{document}